\documentclass[journal=jctcce]{achemso}
\usepackage{graphicx} % Required for inserting images
\usepackage{amsmath}
\usepackage{amssymb}
\usepackage{mathtools}
\newcommand{\erf}{\mbox{erf}}

\title{Range separation of the interaction potential in intermolecular and intramolecular symmetry-adapted perturbation theory}
\author{Du Luu}
\affiliation{Department of Chemistry and Biochemistry, Auburn University,
Auburn, AL 36849, United States}
\author{Clemence Corminboeuf}
\affiliation{Laboratory for Computational Molecular Design, Institute of Chemical Sciences and Engineering, \'Ecole Polytechnique F\'ed\'erale de
Lausanne, CH-1015 Lausanne, Switzerland}
\author{Konrad Patkowski}
\affiliation{Department of Chemistry and Biochemistry, Auburn University,
Auburn, AL 36849, United States}
\email{patkowsk@auburn.edu}

\date{\today}

\begin{document}

\begin{abstract}
Symmetry-adapted perturbation theory (SAPT) is a popular and versatile tool to compute and decompose noncovalent interaction energies between molecules.    
The intramolecular SAPT (ISAPT) variant provides a similar energy decomposition between two nonbonded fragments of the same molecule, covalently connected by a third fragment.
In this work, we explore an alternative approach where the noncovalent interaction is singled out by a range separation of the Coulomb potential.
We investigate two common splittings of the $1/r$ potential into long-range and short-range parts based on the Gaussian and error functions, and approximate either the entire intermolecular/interfragment interaction or only its attractive terms by the long-range contribution.
These range separation schemes are tested for a number of intermolecular and intramolecular complexes.
We find that the energy corrections from range-separated SAPT or ISAPT are in reasonable agreement with complete SAPT/ISAPT data.
This result should be contrasted with the inability of the long-range multipole expansion to describe crucial short-range charge penetration and exchange effects; it shows that the long-range interaction potential does not just recover the asymptotic interaction energy but also provides a useful account of short-range terms.
The best consistency is attained for the error-function separation applied to all interaction terms, both attractive and repulsive.
This study is the first step towards a fragmentation-free decomposition of intramolecular nonbonded energy.
\end{abstract}

\maketitle

\section{Introduction}
Symmetry-adapted perturbation theory (SAPT) \cite{Jeziorski:94, Szalewicz:12, Szalewicz:05a, Hohenstein:12, Patkowski:20} is a popular method for studying intermolecular interactions. The key strength of SAPT is its ability to decompose the interaction energy into physically meaningful components, including electrostatics, induction, dispersion, and their exchange counterparts. 
The SAPT approaches based on a M{\o}ller-Plesset \cite{Rybak:91,Parker:14} or density functional theory (DFT) \cite{Hesselmann:05,Misquitta:05a} description of intramolecular electron correlation are well developed, efficiently implemented workhorses for the ab initio treatment of intermolecular forces.
Many specialized SAPT variants have been proposed and are the subject of ongoing research, targeting, for example, the interactions between multireference molecules \cite{Hapka:19a} and spin splittings due to exchange interaction \cite{Patkowski:18} --- see Ref.~\citenum{Korona:23} for a recent review of other developments.

Notably, some of the new SAPT variants allow the decomposition of nonbonded interactions between fragments of the same molecule \cite{Gonthier:14,Pastorczak:15,Meitei:17,Parrish:15,Luu:23}.
Understanding noncovalent intramolecular interactions is crucial for conformational analysis, chemical reactivity, and molecular design. 
The intramolecular SAPT (ISAPT) method \cite{Parrish:15} relies on a fragmentation of the molecule into two noncovalently interacting fragments A and B embedded into a covalently bonded linker C.
A recent refinement of the fragmentation algorithm has led to consistently meaningful ISAPT interaction energy contributions \cite{Luu:23}. 
However, the total ISAPT interaction energy lacks the contributions arising from the linker (A-C, B-C, and C-C) as well as the ones internal to any single fragment (A-A and B-B). 
Thus, the currently available ISAPT approach provides highly useful insights into the interaction between specific nonbonded fragments, but it does not rise to the challenge of completely describing the noncovalent interaction within a molecule. 
This limits the utility of ISAPT for systems such as hexaphenylethane and its derivatives \cite{Schreiner:11} which exhibit significant stabilization from intramolecular dispersion involving all six bulky groups rather than any individual pair of single-bonded fragments.

In the development of effective computational strategies, the range separation technique is one of versatile tools to delineate the interaction range and to improve convergence of the interaction energy.
The key idea is the splitting of a Coulomb potential $1/r$ into its long-range and short-range parts that are treated differently.
Density functionals incorporating range separation and long-range Hartree-Fock (HF) exchange avoid some common issues such as excessive electron delocalization and the resulting overstabilization of charge-transfer states \cite{Leininger:97,Iikura:01,Vydrov:06}, and they are also useful in the context of SAPT(DFT) as one may avoid an explicit asymptotic correction \cite{Lao:14, Gray:21, Xie:22}. 
On the other hand, density functionals with short-range HF exchange only improve the computational efficiency for periodic systems \cite{Heyd:03}.
Another application of the range separation technique occurs in the regularized SAPT method, addressing the challenges associated with the divergence of the perturbation series \cite{Patkowski:01a,Adams:02,Patkowski:04} as well as allowing a decomposition of induction energy into polarization and charge-transfer contributions \cite{Misquitta:13}. 
Several specific forms of partitioning the interaction potential into short-range and long-range components are available, with the Gaussian and error-function based splittings being the most popular (one should also mention the hybrid ``erfgau'' partitioning \cite{Toulouse:04}). 
The strong point of both splittings is the relatively easy evaluation of the resulting one- and two-electron integrals.

One can expect the long-ranged Coulomb potential to account for nearly all aspects the non-covalent interaction energy including its complete long-range expansion and some (hopefully good) approximation to the short-range overlap effects. 
In this work, we systematically study the convergence behavior of intermolecular and intramolecular SAPT energy contributions using only the long-ranged potential in the interaction operator. 
We investigate a diverse collection of range-separated SAPT methods, including one-electron or both one- and two-electron long-ranged integrals evaluated with Gaussian and error-functions splittings. 
We observe that, for a broad range of the separation parameter values, the purely long-range SAPT and ISAPT contributions provide reliable approximations to the complete non-range separated quantities, indicating that the noncovalent interaction can be singled out via range separation rather than molecular fragmentation.
Our proof of concept study, inspired by a similar long-range vs. short-range splitting in the context of supermolecular interaction energies \cite{Chen:16}, suggests range separation as a promising first step towards developing a fragmentation-free ISAPT algorithm. 

All range separation-based SAPT and ISAPT methods proposed here are implemented into the Psi4 software \cite{Smith:20} and available on GitHub: \verb+https://github.com/dluu12/psi4/tree/rsep+.

\section{Theory}

The inner workings of the intermolecular symmetry-adapted perturbation theory (SAPT) are described in a number of reviews; see, for example, Refs.~\citenum{Jeziorski:94} and \citenum{Patkowski:20}.
In this work, we will examine the simplest level of SAPT that entirely neglects intramonomer electron correlation and is usually termed SAPT0.
The SAPT0 interaction energy is decomposed as
\begin{equation}
\label{eq:sapt0}
E^{\rm SAPT0}_{\rm int} = \left(E^{(10)}_{\rm elst}\right)+ \left(E^{(10)}_{\rm exch}\right)+ \left(E^{(20)}_{\rm ind,resp}+ E^{(20)}_{\rm exch-ind,resp}+ \delta E^{\rm HF}_{\rm int}\right)+
\left(E^{(20)}_{\rm disp}+ E^{(20)}_{\rm exch-disp}\right)
\end{equation}
The SAPT0 corrections $E^{(kl)}$, grouped by parentheses into the total electrostatic, exchange, induction, and dispersion terms, are labeled by their order $k$ in the intermonomer interaction operator and their order $l$ in the intramonomer correlation operator.
The additional subscript ``resp'' denotes the coupled perturbed Hartree-Fock response (relaxation) of monomer orbitals in the electric field of the interacting partner, and $\delta E^{\rm HF}_{\rm int}$ is an estimate of higher-order induction and exchange-induction effects from the supermolecular HF approach.
While many higher-level intermolecular SAPT variants are available, employing different strategies to capture intramonomer correlation \cite{Rybak:91,Misquitta:05a,Hesselmann:05,Korona:09a,Parker:14}, intramolecular SAPT (ISAPT) is only available at the SAPT0 level of perturbation theory \cite{Parrish:15,Luu:23}.

In any SAPT variant, the perturbation operator $V$ is the intermonomer interaction operator, containing Coulombic attraction/repulsion terms between all particles (nuclei and electrons) of A with those of B. 
In this work, we will examine the effects of separating $V$ into its long-range and short-range part by a partitioning of the Coulomb potential,
\begin{equation}
\frac{1}{r}=v_{\rm l}(r)+v_{\rm s}(r)    
\end{equation}
using either the Gaussian function splitting
\begin{equation}\label{eq:gaureg}
v_{\rm l}(r) = \frac{1-e^{-\eta r^2}}{r} \;\;\;\;\;\;
v_{\rm s}(r) = \frac{e^{-\eta r^2}}{r}
\end{equation}
or the error-function (erf) splitting
\begin{equation}\label{eq:erfreg}
v_{\rm l}(r) = \frac{\erf(\omega r)}{r} \;\;\;\;\;\;
v_{\rm s}(r) = \frac{1-\erf(\omega r)}{r}
\end{equation}
The parameters $\omega$ (in units of (distance)$^{-1}$) and $\eta$ (in units of (distance)$^{-2}$) define the width of the range separation.
The erf splitting of Eq.~(\ref{eq:erfreg}) is ubiquitous in range-separated density functional theory (DFT) \cite{Iikura:01,Vydrov:06}.
The Gaussian splitting of Eq.~(\ref{eq:gaureg}) is less prevalent but has been used before in the context of SAPT to regularize the attractive singularities in the Coulomb potential \cite{Patkowski:04,Misquitta:13} (more on that process below).
Overall, we expect that $v_{\rm l}(r)$ will play a much bigger role in any SAPT corrections than $v_{\rm s}(r)$: the long-range potential should be able to recover the entire asymptotic expansion of interaction energy in inverse powers of the intermolecular distance $R$ as well as at least some of the overlap-dependent, exponentially vanishing terms.
Therefore, we will examine the role of the approximation $1/r\approx v_{\rm l}(r)$, that is, the neglect of the short-range part of the interaction operator $V$, on various SAPT corrections.
It is possible to do the opposite and approximate $1/r$ by its short-range part $v_{\rm s}(r)$ (in fact, both of these approximations have been tested before for the supermolecular interaction energy curves of rare gas dimers \cite{Chen:16}); however, we will not pursue this direction as the long-range interaction is obviously crucial to obtain a reasonable, asymptotically meaningful interaction energy.

The short-range Coulomb interaction can be neglected in different parts of the interaction operator.
For example, only the attractive, one-electron part of $V$ can be replaced by its long-range component:
\begin{equation}\label{eq:reg1el}
V_{\rm 1e}=-\sum_{A\in\mathbf{A}}\sum_{j\in\mathbf{B}} Z_A v_{\rm l}(r_{Aj}) -
\sum_{B\in\mathbf{B}}\sum_{i\in\mathbf{A}} Z_B v_{\rm l}(r_{Bi}) +
\sum_{i\in\mathbf{A}}\sum_{j\in\mathbf{B}} \frac{1}{r_{ij}} +
\sum_{A\in\mathbf{A}}\sum_{B\in\mathbf{B}} \frac{Z_A Z_B}{r_{AB}}
\end{equation}
The rationale behind Eq.~(\ref{eq:reg1el}) is that the attractive singularities in $V$ are more consequential than the repulsive ones, as the electrons from one monomer are pulled into the Coulomb wells around the other monomer's nuclei (note that the perturbation wavefunctions exist in a space spanned by products of monomer wavefunctions, and the permutation symmetry for electron interchanges between monomers is not enforced).
This overpolarization of monomers resulting from the attractive singularities is the reason for the overall divergence of the SAPT series \cite{Adams:91,Patkowski:04}, which manifests itself in practice by the induction and exchange-induction energies increasing in magnitude with increasing order of SAPT (note that the dispersion energy does not depend on one-electron integrals so it is not directly affected).
Accordingly, the removal {\it (regularization)} of attractive singularities in Eq.~(\ref{eq:reg1el}) was proposed to make the SAPT expansion convergent \cite{Patkowski:01a,Adams:02,Patkowski:04}.
While an outright neglect of $v_{\rm s}(r)$ in Eq.~(\ref{eq:reg1el}) obviously alters the limit of the perturbation series, this operator can be brought back by the so-called {\em strong symmetry forcing} \cite{Jeziorski:94}. 
The resulting perturbation expansion was shown for a model Li--H system to converge to the full configuration interaction (FCI) interaction energy  \cite{Patkowski:04}.
A different application of regularized SAPT based on Eq.~(\ref{eq:reg1el}) was proposed by Misquitta \cite{Misquitta:13} and involves the (approximate) partitioning of induction energy into polarization and charge transfer (charge delocalization) contributions.
The charge-transfer term involves the tunneling of electrons into the other monomer's attractive Coulomb wells, so it is suppressed by regularization of Eq.~(\ref{eq:reg1el}) while the polarization term remains.

Alternatively, the short-range part of the Coulomb potential can be neglected in all terms in $V$, attractive and repulsive:
\begin{equation}\label{eq:regfull}
V_{\rm 1e2e}=-\sum_{A\in\mathbf{A}}\sum_{j\in\mathbf{B}} Z_A v_{\rm l}(r_{Aj}) -
\sum_{B\in\mathbf{B}}\sum_{i\in\mathbf{A}} Z_B v_{\rm l}(r_{Bi}) +
\sum_{i\in\mathbf{A}}\sum_{j\in\mathbf{B}} v_{\rm l}(r_{ij}) +
\sum_{A\in\mathbf{A}}\sum_{B\in\mathbf{B}} Z_A Z_B v_{\rm l}(r_{AB})
\end{equation}
This form of perturbation might potentially be preferable over $V_{\rm 1e}$ of Eq.~(\ref{eq:reg1el}) as the attractive and repulsive terms are treated on an equal footing.
It is well known that the SAPT terms, most notably the electrostatic energy, are a result of a delicate balance between the attractive and repulsive Coulomb forces, and this balance could be disturbed when only the attractive forces are attenuated at short range.
The potential $V_{\rm 1e2e}$ was briefly tested in the context of regularized SAPT for the H--H complex \cite{Patkowski:01a}, leading to similar results as $V_{\rm 1e}$, however, more testing for larger complexes is warranted.

Let us point out some important differences in the goals and implementation of the range separated SAPT in this work and the regularized SAPT approach of Refs.~\citenum{Patkowski:01a} and \citenum{Patkowski:04}, specifically, the regularized symmetrized Rayleigh-Schr\"{o}dinger (R-SRS) theory adapted in Ref.~\citenum{Misquitta:13}.
On the technical side, we will be neglecting the attractive (Eq.~(\ref{eq:reg1el})) or all (Eq.~(\ref{eq:regfull})) short-range terms completely, replacing $V$ by either $V_{\rm 1e}$ or $V_{\rm 1e2e}$ throughout the entire SAPT calculation.
On the other hand, the R-SRS perturbation theory employs $V_{\rm 1e}/V_{\rm 1e2e}$ in the calculation of wavefunction corrections (where convergence needs to be enforced) but switches to the entire $V$ in the subsequent computation of the SRS energy terms from the wavefunction corrections.
The R-SRS approach is specifically geared at computing the induction and exchange-induction terms, where it can be used to achieve a convergent series (Ref.~\citenum{Patkowski:04}) or to separate the polarization and charge-transfer contributions (Ref.~\citenum{Misquitta:13}).
However, the range separation studied here applies to all SAPT0/ISAPT terms except for $\delta E^{\rm HF}_{\rm int}$.
The main goal of this range separation is to check whether the long-range and short-range part of $V$ can be identified with noncovalent and covalent interactions, respectively, so that the neglect of the latter does not appreciably influence the SAPT0/ISAPT corrections.
%and, possibly, the neglect of the former does not adversely effect the embedding in the ISAPT linker fragment.
%We will also examine how the replacement of $V$ by either $V_{\rm 1e}$ or $V_{\rm full}$ affects the molecular electrostatic potential (ESP) at noncovalent distances. {\bf KP: we could do it but we don't absolutely have to.}

In the ISAPT method \cite{Parrish:15}, the molecule is partitioned into noncovalently interacting fragments A and B and a linker C that connects to A and B by single covalent bonds. 
Then, the occupied orbitals and density matrices of noninteracting A and B are obtaining by solving the Hartree-Fock equations for each fragment embedded in the Fock matrix of linker C.
Those orbitals and density matrices are then used in standard intermolecular SAPT0 expressions to compute different noncovalent intramolecular energy contributions.
It should be noted that the ISAPT accuracy is sensitive to the specific algorithm used to partition the electron density into fragments, and the original partitioning of Ref.~\citenum{Parrish:15} often leads to meaningless results due to unphysical dipole moments emerging at the interfragment boundaries.
To remedy this problem, we have recently proposed \cite{Luu:23} a number of alternative partitionings aimed at reducing those spurious dipoles.
Numerical tests indicated that one of the new algorithms, which assigns one electron from the interfragment bonding orbital to A (B) and places it onto an orbital obtained by a projection onto the space of intrinsic atomic orbitals (IAOs) \cite{Knizia:13} of a fragment, exhibits both meaningful ISAPT partitioning for all fragmentation patterns and smooth basis set convergence.
This algorithm, termed SIAO1 as it {\bf S}plits link orbitals in the {\bf IAO} space by {\bf 1} iteration of a self-consistent algorithm, was recommended in Ref.~\citenum{Luu:23} and will be employed in this work.
Here, we examine how the replacement of the full interfragment interaction operator $V$ by either $V_{\rm 1e}$ or $V_{\rm 1e2e}$ affects the ISAPT energy contributions. As noncovalently interacting fragments are typically closer together than noncovalently interacting separate molecules, we expect ISAPT to be a harder test case than SAPT0, placing more severe restrictions on the range separation parameter $\eta$ ($\omega$). 

\section{Results and Discussion}

The range-separated SAPT approach requires new one- and two-electron integrals involving either $v_{\rm l}(r)$ or $v_{\rm s}(r)$ (integrals of the other type are then obtained by subtraction from the standard integrals involving $1/r$). 
Specifically, in the Gaussian splitting approach of Eq.~(\ref{eq:gaureg}) we implemented the $v_{\rm s}(r)$ integrals, and for the error-function splitting of Eq.~(\ref{eq:erfreg}) we programmed the $v_{\rm l}(r)$ integrals.
Both two-electron integral types were already available in the {\sc psi4} code used for our development \cite{Smith:20}: the error-function one as a key building block of range-separated density functional integrals \cite{Iikura:01,Vydrov:06} and the Gaussian one as a standard integral in the explicitly correlated F12 theory \cite{Hattig:12,Kong:12}.
On the other hand, the corresponding one-electron integrals were derived and programmed by us utilizing the McMurchie-Davidson angular momentum recursion \cite{McMurchie:81} -- see Appendix for formulas.
The resulting integrals were compared to our older unpublished code utilizing the Obara-Saika recursion \cite{Obara:86} to verify the correctness of our implementation.
The integrals were implemented into a development version of {\sc psi4} and are accessible from both the C$++$ and Python layers.

All the range-separated intermolecular SAPT approaches utilizing the $v_{\rm l}(r)$ Coulombic potential, partitioned with either the Gaussian function (Eq.~(\ref{eq:gaureg})) or the error function (Eq.~(\ref{eq:erfreg})), were examined and analyzed using the Psi4NumPy framework \cite{Smith:18}. 
Four noncovalent systems were chosen to study the behavior of the interaction energy as a function of the intermolecular distance $R$: the methane dimer, water dimer, the ammonia-chlorine monofluoride complex, and the water-fluoride ion one. 
The geometry optimizations for the $\text H_{2}\text O \cdots \text H_{2} \text O$ , $\text H_{2}\text O \cdots \text {F}^{-}$, and $\text {NH}_{3} \cdots \text {ClF}$ systems are performed at the MP2 level of theory using the aug-cc-pVDZ basis set. 
The structure of the $\text {CH}_{4}\cdots \text {CH}_{4}$ complex is taken from the S22 \cite{Jurecka:06} database. The single-point interaction energies as the function of $R$, the distance between the centers of masses of the two monomers, with multiple range separation parameters are computed using the SAPT0/jun-cc-pVDZ method. 

The relationship between the separation parameters $\eta$ (for the Gaussian function splitting) and $\omega$ (for the error function splitting) needs to be carefully described. Since the unit of $\eta$ is $r^{-2}$ and of $\omega$ is $r^{-1}$, a specific value of $\eta$ is expected to correspond to $k\omega^2$ for some constant $k$. Therefore, to compare the behavior of the two range separations, we will investigate the values of $\omega$ equal to the square roots of the corresponding values of $\eta$.

\begin{figure}[!htb]
\begin{minipage}{.48\textwidth}
\centering
\includegraphics[width=1\textwidth]{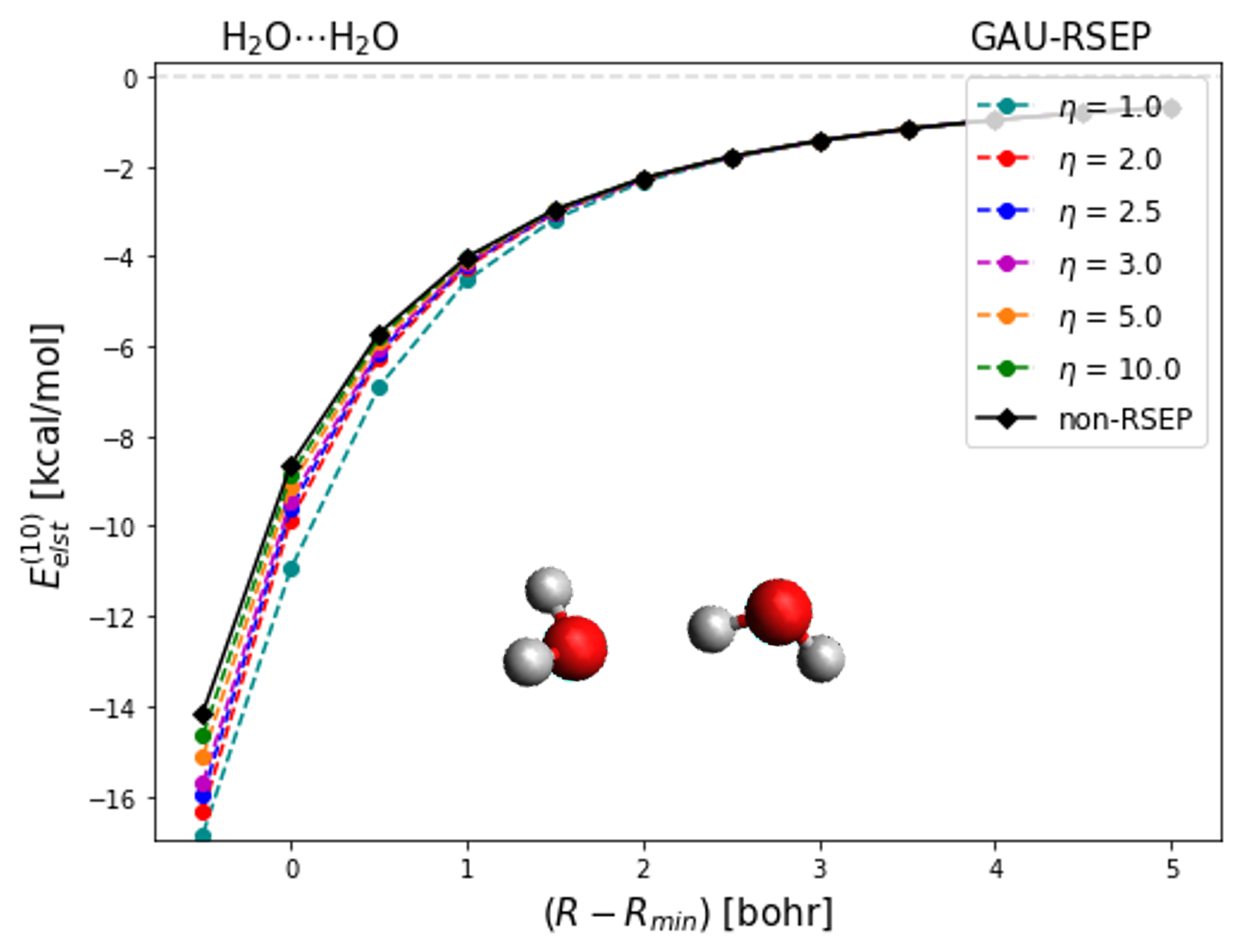}
\end{minipage}
\begin{minipage}{.48\textwidth}
\centering
\includegraphics[width=1\textwidth]{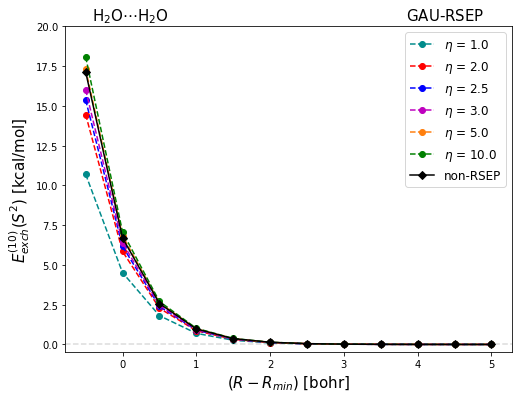}
\end{minipage}
\begin{minipage}{.48\textwidth}
\centering
\includegraphics[width=1\textwidth]{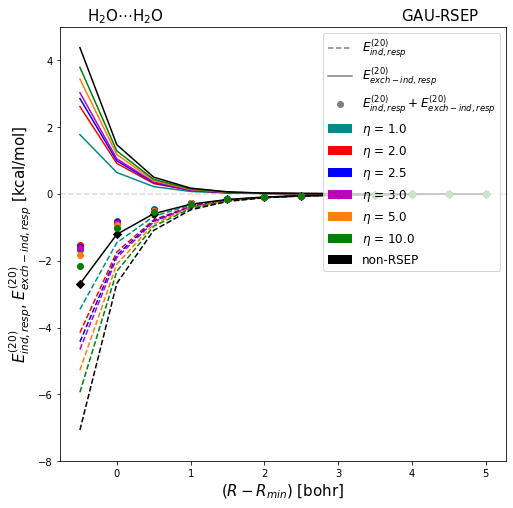}
\end{minipage}
\begin{minipage}{.48\textwidth}
\centering
\includegraphics[width=1\textwidth]{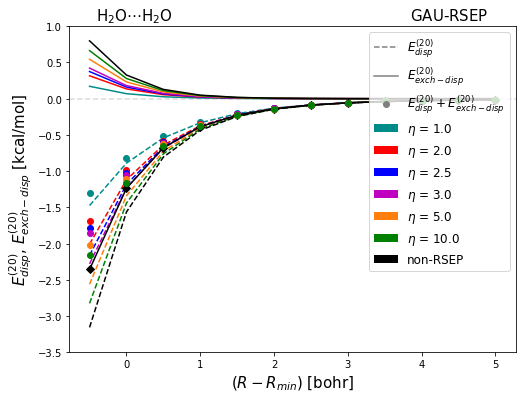}
\end{minipage}
\begin{minipage}{.48\textwidth}
\centering
\includegraphics[width=1\textwidth]{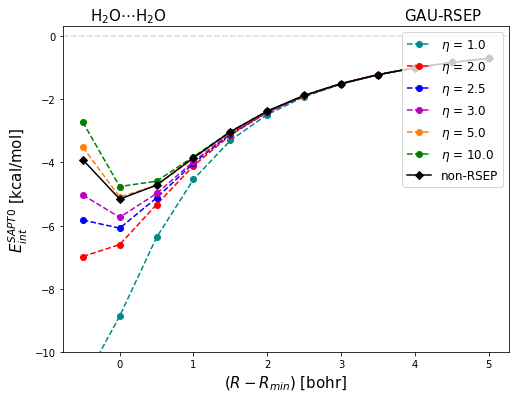}
\end{minipage}
\caption{The SAPT0(RSEP/gau) interaction energy components in the water dimer computed with various values of $\eta$ (bohr$^{-2}$) in the jun-cc-pVDZ basis set. }
\label{fig:h2o-gau-rsep-plot}
\end{figure}

\begin{figure}[!htb]
\begin{minipage}{.48\textwidth}
\centering
\includegraphics[width=1\textwidth]{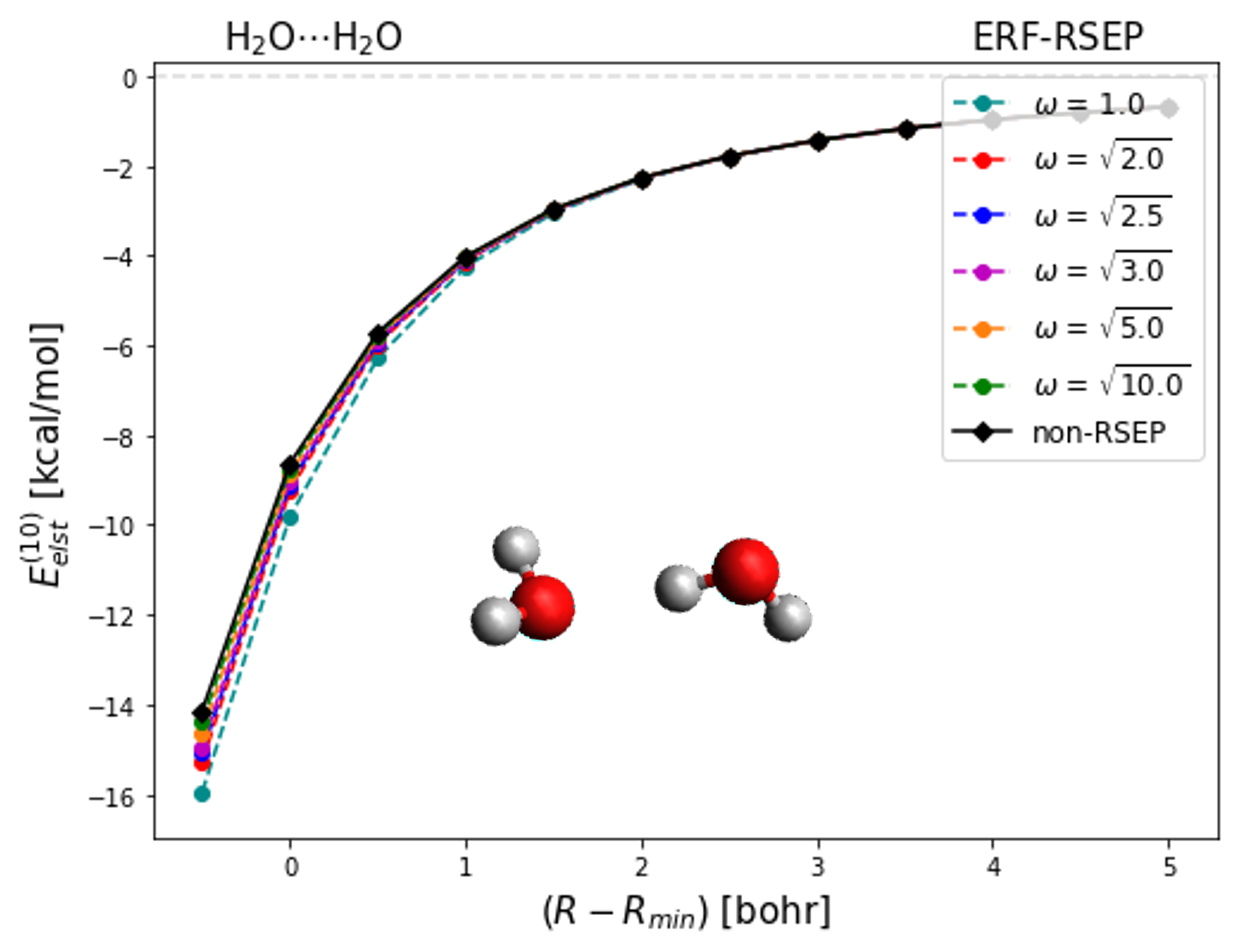}
\end{minipage}
\begin{minipage}{.48\textwidth}
\centering
\includegraphics[width=1\textwidth]{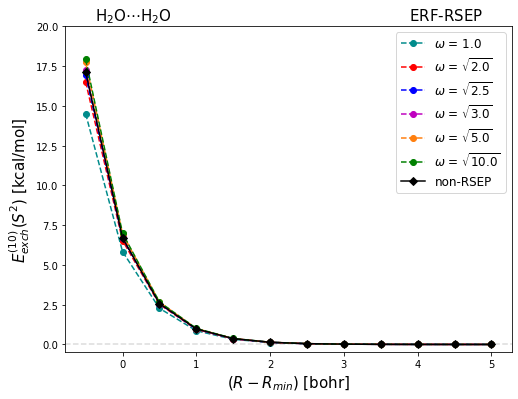}
\end{minipage}
\begin{minipage}{.48\textwidth}
\centering
\includegraphics[width=1\textwidth]{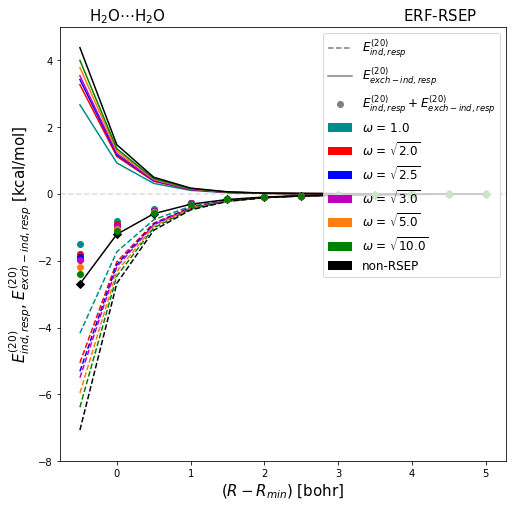}
\end{minipage}
\begin{minipage}{.48\textwidth}
\centering
\includegraphics[width=1\textwidth]{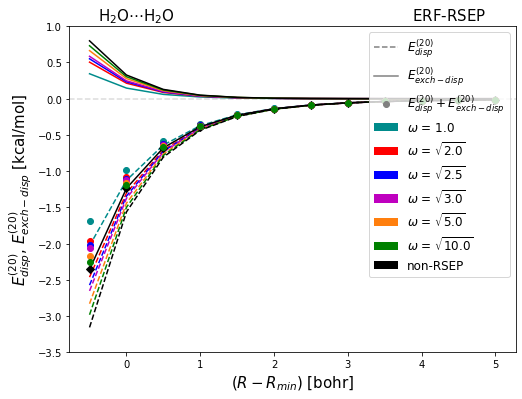}
\end{minipage}
\begin{minipage}{.48\textwidth}
\centering
\includegraphics[width=1\textwidth]{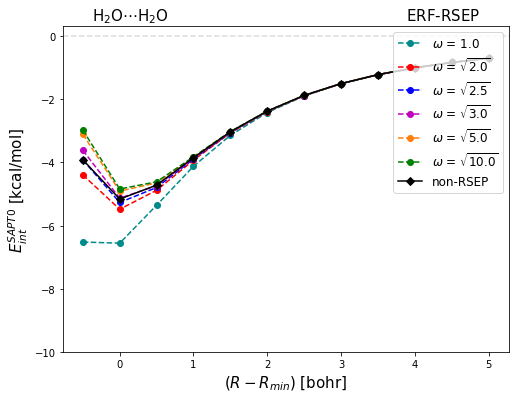}
\end{minipage}
\caption{The SAPT0(RSEP/erf) interaction energy components in the water dimer computed with various values of $\omega$ (bohr$^{-1}$) in the jun-cc-pVDZ  basis set.  }
\label{fig:h2o-erf-rsep-plot}
\end{figure}

\begin{figure}[!htb]
\begin{minipage}{.48\textwidth}
\centering
\includegraphics[width=1\textwidth]{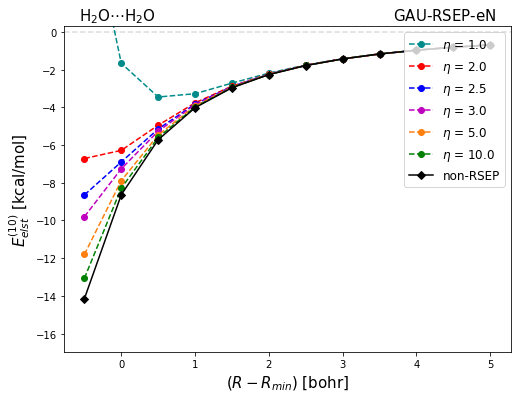}
\end{minipage}
\begin{minipage}{.48\textwidth}
\centering
\includegraphics[width=1\textwidth]{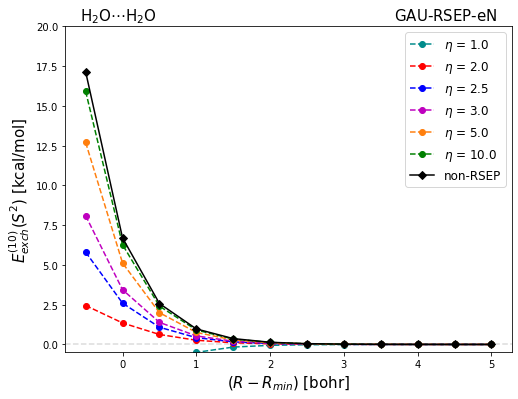}
\end{minipage}
\begin{minipage}{.48\textwidth}
\centering
\includegraphics[width=1\textwidth]{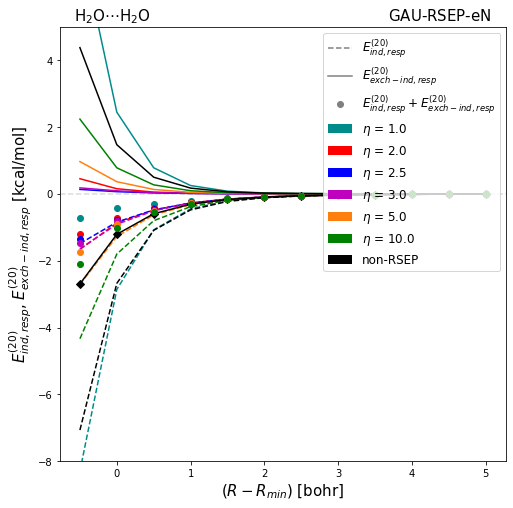}
\end{minipage}
\begin{minipage}{.48\textwidth}
\centering
\includegraphics[width=1\textwidth]{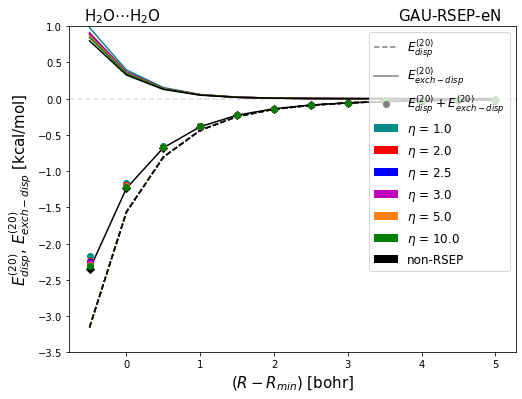}
\end{minipage}
\begin{minipage}{.48\textwidth}
\centering
\includegraphics[width=1\textwidth]{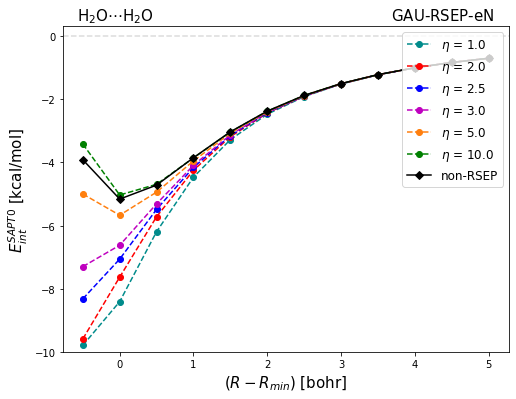}
\end{minipage}
\caption{The SAPT0(RSEP-eN/gau) interaction energy components in the water dimer computed with various values of $\eta$ (bohr$^{-2}$) in the jun-cc-pVDZ basis set.  }
\label{fig:h2o-gau-rsep-en-plot}
\end{figure}

\begin{figure}[!htb]
\begin{minipage}{.48\textwidth}
\centering
\includegraphics[width=1\textwidth]{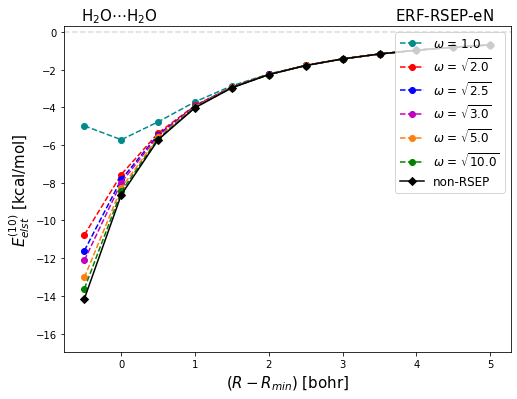}
\end{minipage}
\begin{minipage}{.48\textwidth}
\centering
\includegraphics[width=1\textwidth]{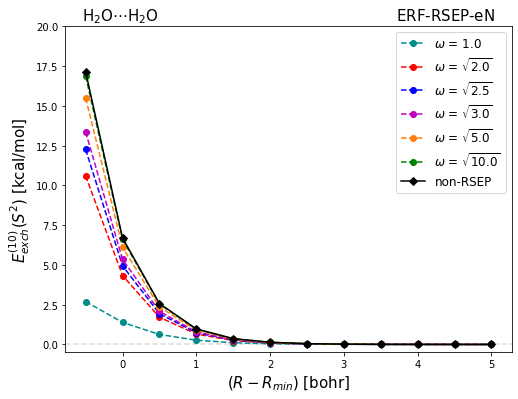}
\end{minipage}
\begin{minipage}{.48\textwidth}
\centering
\includegraphics[width=1\textwidth]{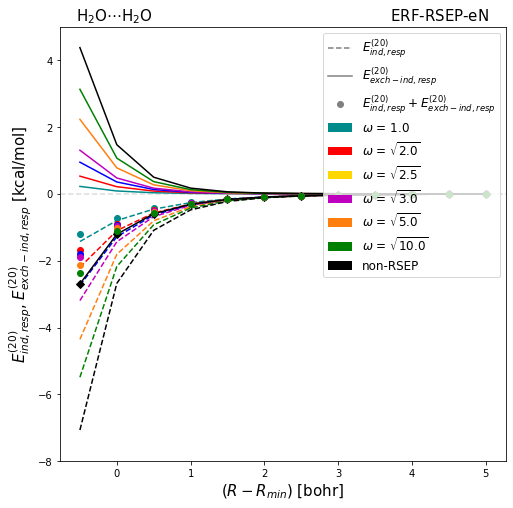}
\end{minipage}
\begin{minipage}{.48\textwidth}
\centering
\includegraphics[width=1\textwidth]{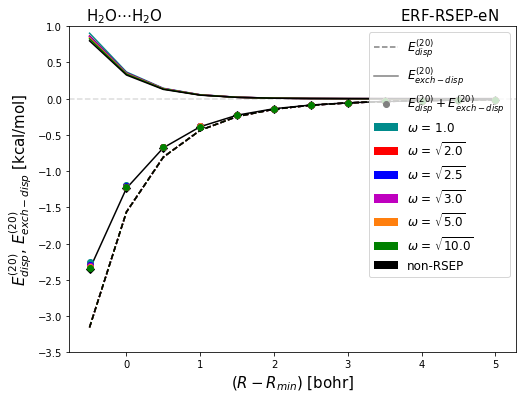}
\end{minipage}
\begin{minipage}{.48\textwidth}
\centering
\includegraphics[width=1\textwidth]{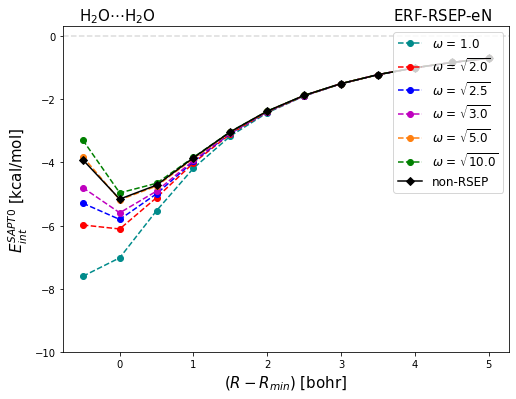}
\end{minipage}
\caption{The SAPT0(RSEP-eN/erf) interaction energy components in the water dimer computed with various values of $\omega$ (bohr$^{-1}$) in the jun-cc-pVDZ basis set.  }
\label{fig:h2o-erf-rsep-en-plot}
\end{figure}

\begin{figure}[!htb]
\begin{minipage}{.48\textwidth}
\centering
\includegraphics[width=1\textwidth]{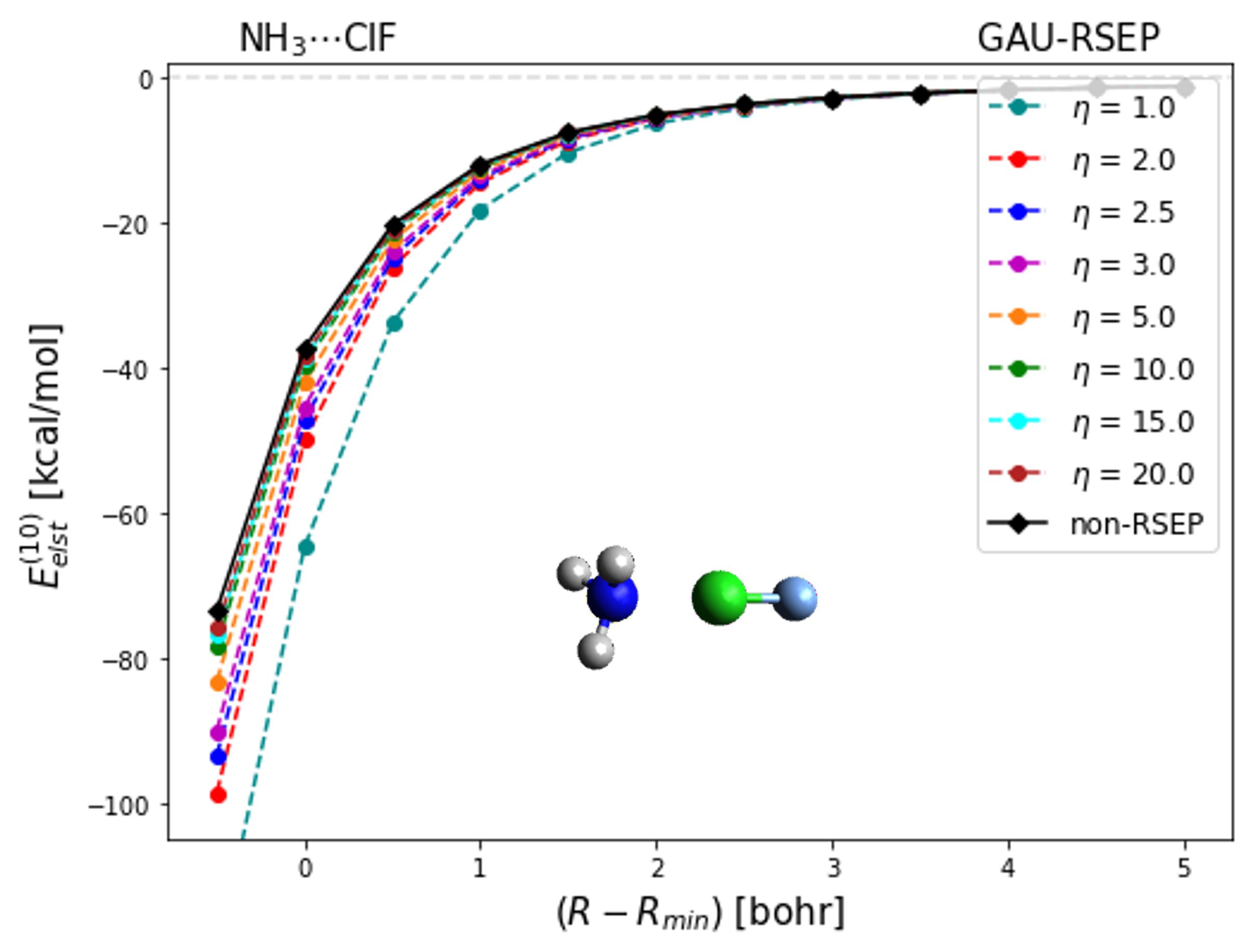}
\end{minipage}
\begin{minipage}{.48\textwidth}
\centering
\includegraphics[width=1\textwidth]{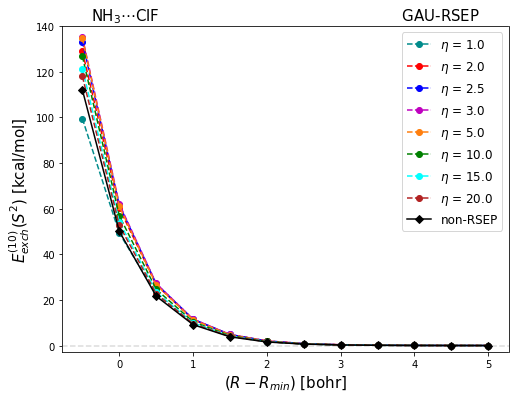}
\end{minipage}
\begin{minipage}{.48\textwidth}
\centering
\includegraphics[width=1\textwidth]{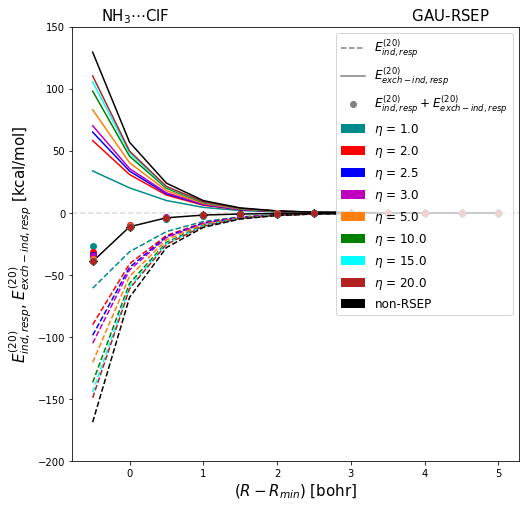}
\end{minipage}
\begin{minipage}{.48\textwidth}
\centering
\includegraphics[width=1\textwidth]{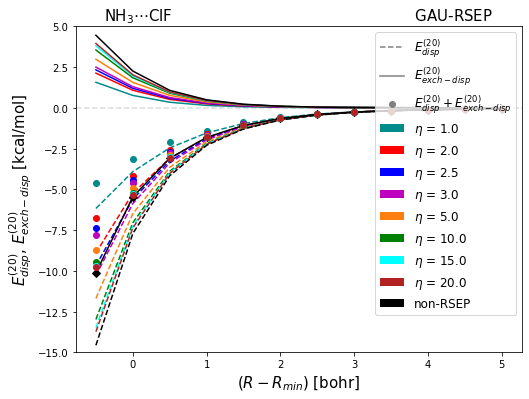}
\end{minipage}
\begin{minipage}{.48\textwidth}
\centering
\includegraphics[width=1\textwidth]{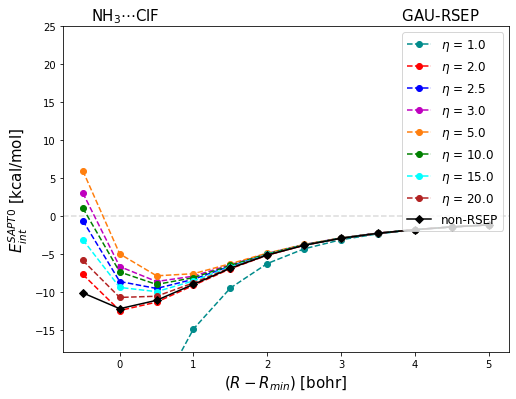}
\end{minipage}
\caption{The SAPT0(RSEP/gau) interaction energy components in the ammonia-chlorine monofluoride complex computed with various values of $\eta$ (bohr$^{-2}$) in the jun-cc-pVDZ basis set. }
\label{fig:nh3clf-gau-rsep-plot}
\end{figure}

\begin{figure}[!htb]
\begin{minipage}{.48\textwidth}
\centering
\includegraphics[width=1\textwidth]{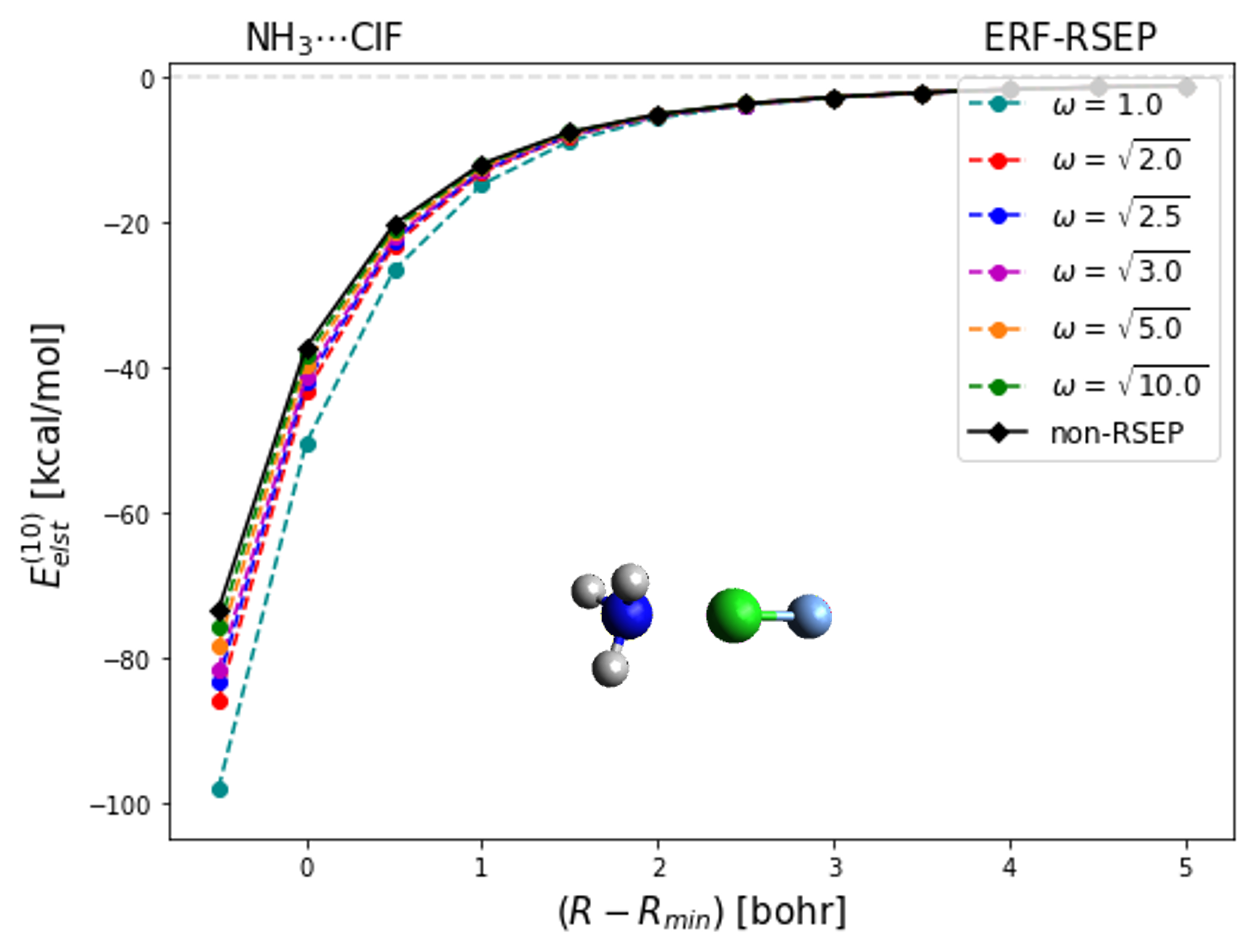}
\end{minipage}
\begin{minipage}{.48\textwidth}
\centering
\includegraphics[width=1\textwidth]{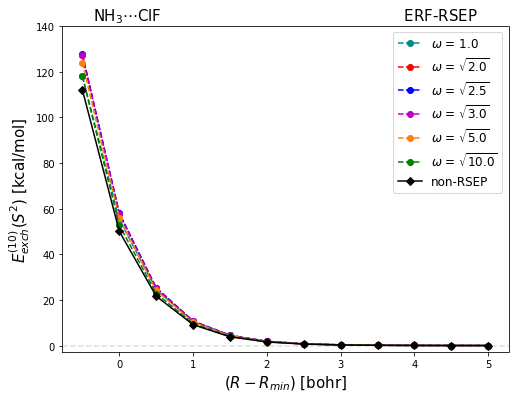}
\end{minipage}
\begin{minipage}{.48\textwidth}
\centering
\includegraphics[width=1\textwidth]{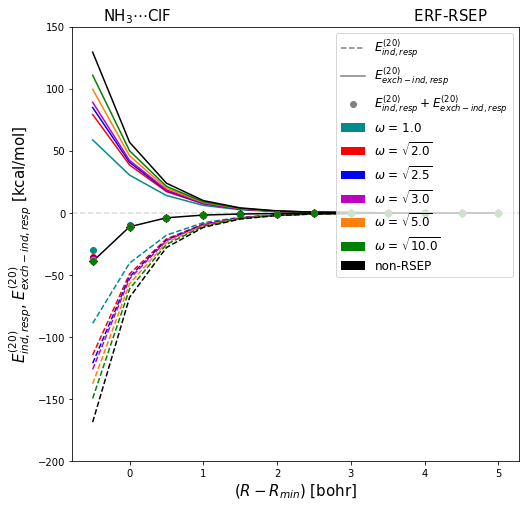}
\end{minipage}
\begin{minipage}{.48\textwidth}
\centering
\includegraphics[width=1\textwidth]{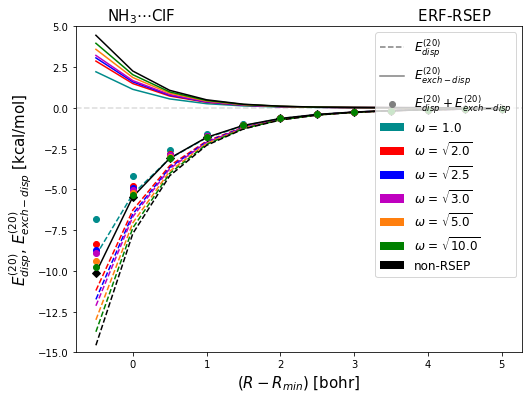}
\end{minipage}
\begin{minipage}{.48\textwidth}
\centering
\includegraphics[width=1\textwidth]{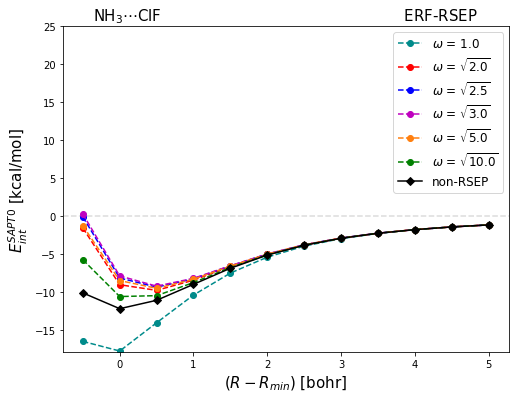}
\end{minipage}
\caption{The SAPT0(RSEP/erf) interaction energy components in the ammonia-chlorine monofluoride complex computed with various values of $\omega$ (bohr$^{-1}$) in the jun-cc-pVDZ  basis set.  }
\label{fig:nh3clf-erf-rsep-plot}
\end{figure}

\begin{figure}[!htb]
\begin{minipage}{.48\textwidth}
\centering
\includegraphics[width=1\textwidth]{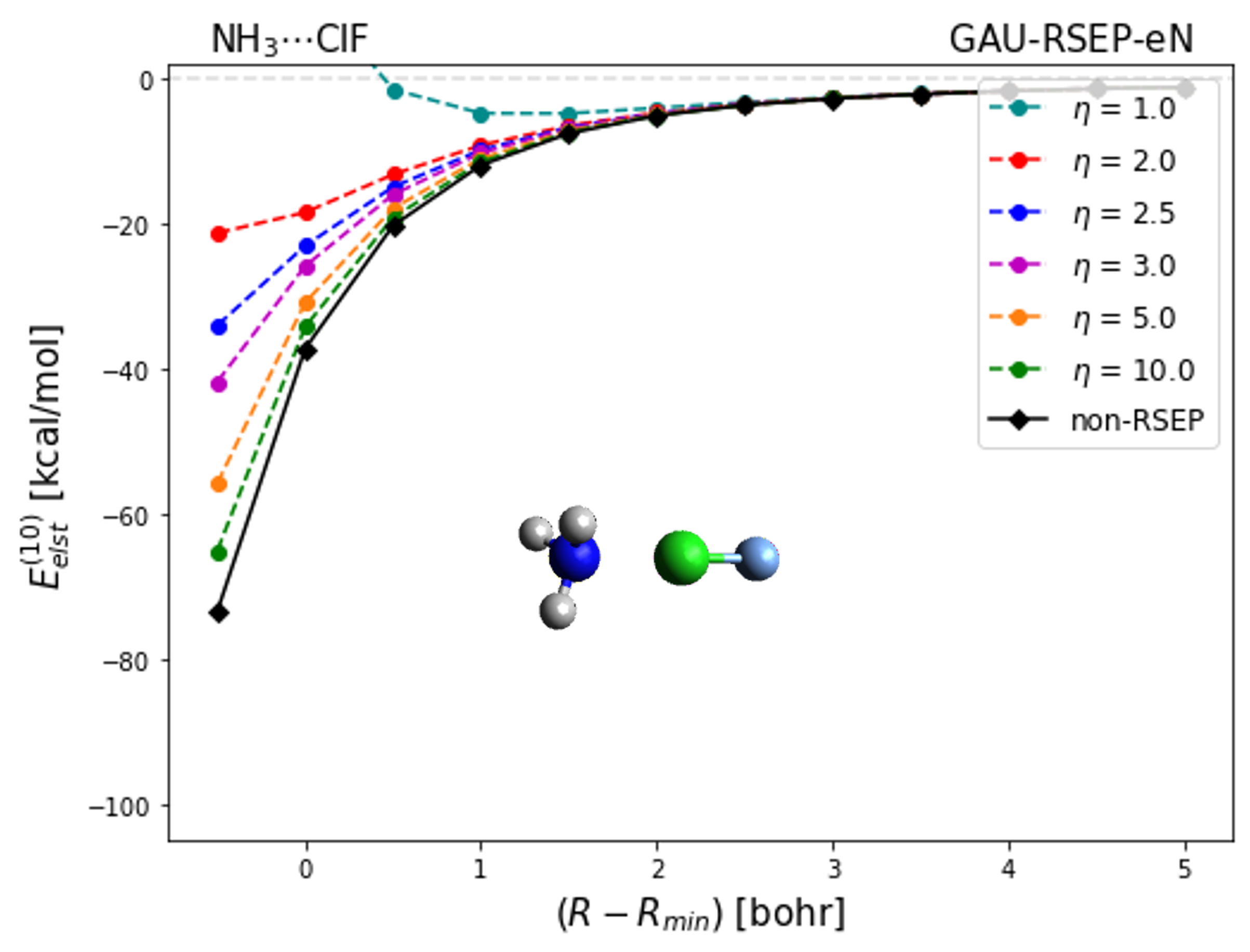}
\end{minipage}
\begin{minipage}{.48\textwidth}
\centering
\includegraphics[width=1\textwidth]{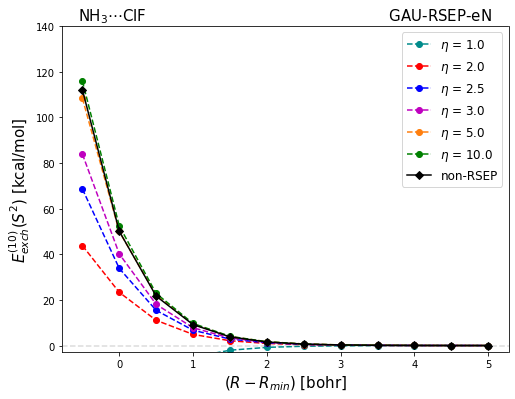}
\end{minipage}
\begin{minipage}{.48\textwidth}
\centering
\includegraphics[width=1\textwidth]{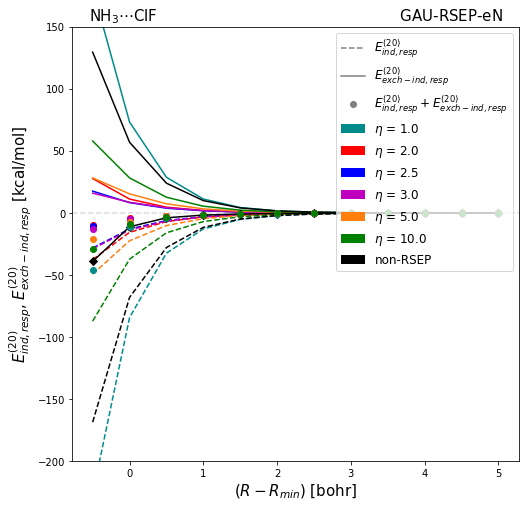}
\end{minipage}
\begin{minipage}{.48\textwidth}
\centering
\includegraphics[width=1\textwidth]{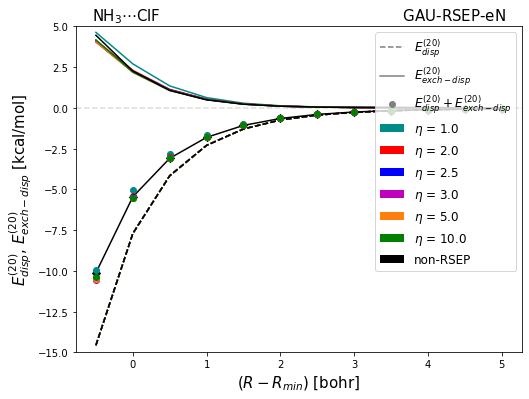}
\end{minipage}
\begin{minipage}{.48\textwidth}
\centering
\includegraphics[width=1\textwidth]{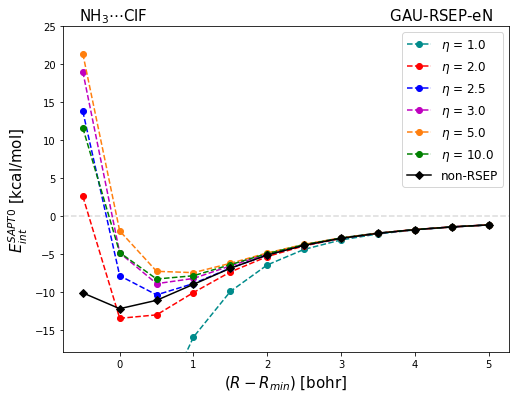}
\end{minipage}
\caption{The SAPT0(RSEP-eN/gau) interaction energy components in the ammonia-chlorine monofluoride complex computed with various values of $\eta$ (bohr$^{-2}$) in the jun-cc-pVDZ basis set.  }
\label{fig:nh3clf-gau-rsep-en-plot}
\end{figure}

\begin{figure}[!htb]
\begin{minipage}{.48\textwidth}
\centering
\includegraphics[width=1\textwidth]{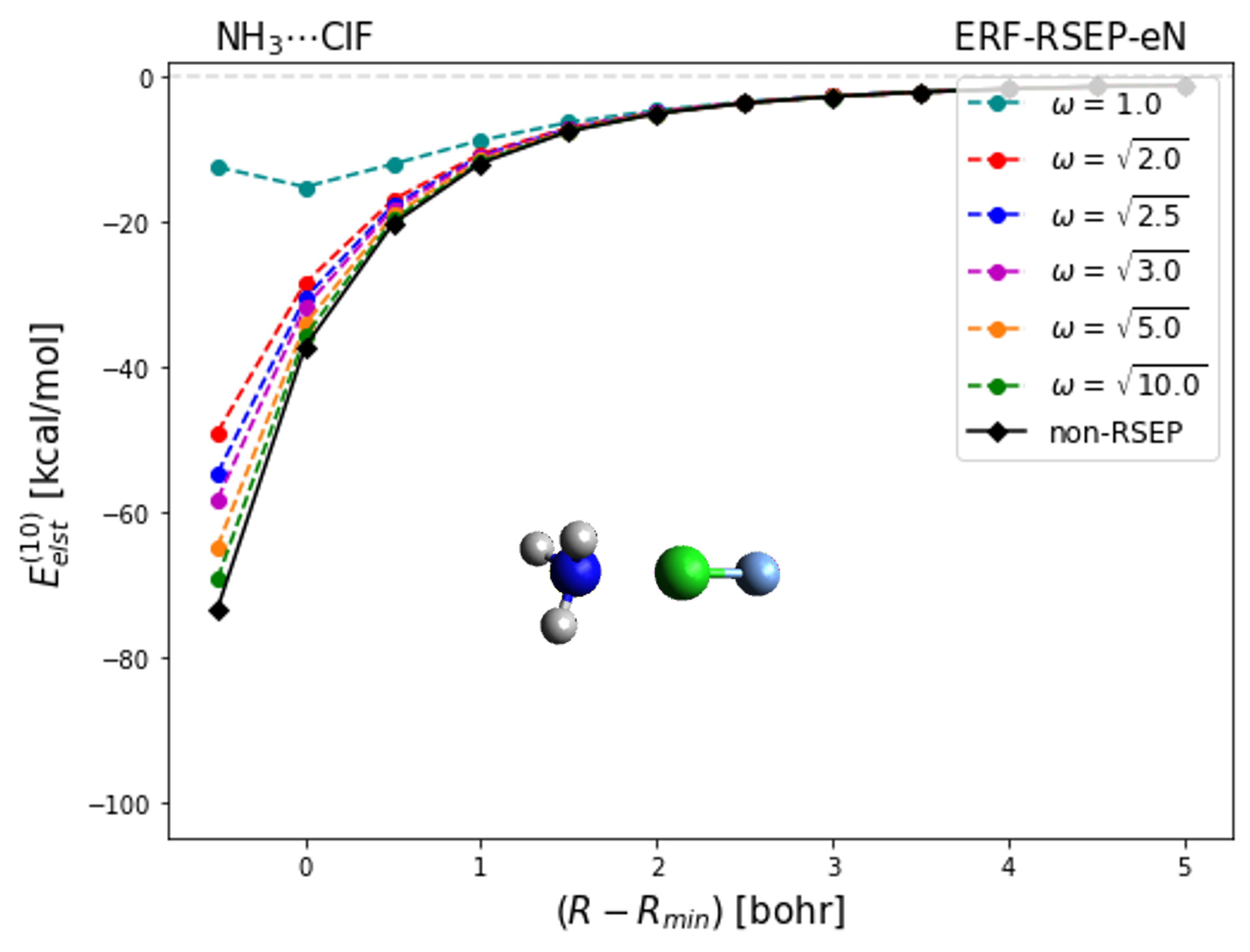}
\end{minipage}
\begin{minipage}{.48\textwidth}
\centering
\includegraphics[width=1\textwidth]{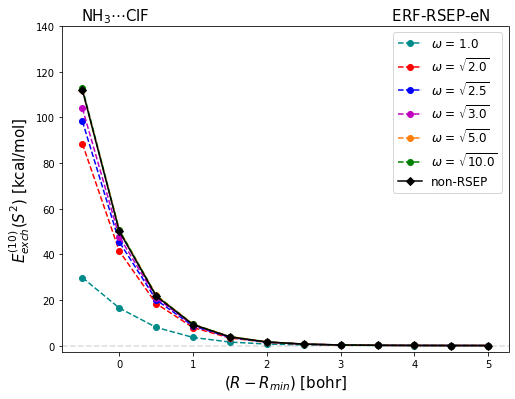}
\end{minipage}
\begin{minipage}{.48\textwidth}
\centering
\includegraphics[width=1\textwidth]{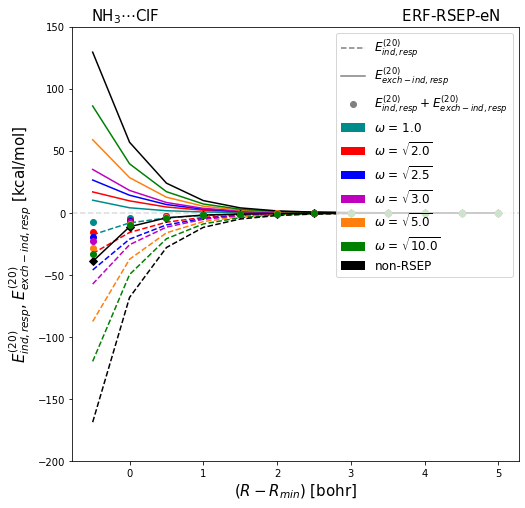}
\end{minipage}
\begin{minipage}{.48\textwidth}
\centering
\includegraphics[width=1\textwidth]{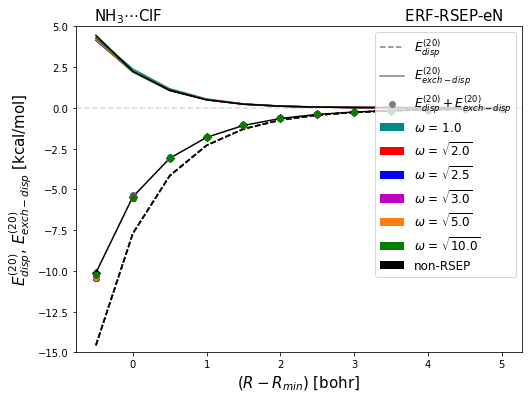}
\end{minipage}
\begin{minipage}{.48\textwidth}
\centering
\includegraphics[width=1\textwidth]{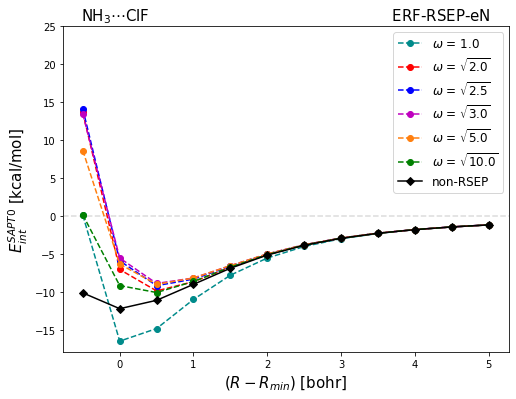}
\end{minipage}
\caption{The SAPT0(RSEP-eN/erf) interaction energy components in the ammonia-chlorine monofluoride complex computed with various values of $\omega$ (bohr$^{-1}$) in the jun-cc-pVDZ basis set.  }
\label{fig:nh3clf-erf-rsep-en-plot}
\end{figure}

Figures \ref{fig:h2o-gau-rsep-plot}--\ref{fig:h2o-erf-rsep-en-plot} present the interaction energies in the water dimer computed with the full range-separated SAPT (Eq.~(\ref{eq:regfull}), denoted RSEP for range separated electrostatic potential) and the one-electron range-separated SAPT (Eq.~(\ref{eq:reg1el}), denoted RSEP-eN). The corresponding figures for the $\text {NH}_{3} \cdots \text {ClF}$ complex are shown in Figs.~\ref{fig:nh3clf-gau-rsep-plot}--\ref{fig:nh3clf-erf-rsep-en-plot}, and those for the two remaining systems are provided in the Supporting Information. As expected, all range-separated variants converge to the full SAPT0 values at large $R$.
More importantly, however, the long-ranged SAPT terms maintain close agreement with the full SAPT numbers at intermediate $R$, down to the van der Waals minimum separation or even slightly closer, except for $\eta=1.0$ bohr$^{-2}$ and $\omega=1.0$ bohr$^{-1}$ where the range separation is clearly too strong.
This is exactly the physically reasonable range of separation parameters -- note that Misquitta suggested $\eta=3.0$ bohr$^{-2}$ to separate polarization and charge transfer within the R-SRS method \cite{Misquitta:13}.

%Overall, in both splitting of Gaussian and Error functions, the full range-separated SAPT (RSEP) method exhibits the strongest regularization on the interaction energy distributions, followed by the R-SRS and the one-electron range-separated SAPT (RSEP-eN) methods. As expected, the RSEP method is implemented with the $v_{l}(r)$ in both attractively and repulsively electronic potentials, which should strongly recover the regular SAPT0 results. The RSEP-eN method is only contributed by the one-electron integral of the Coulombic potential, whereas the R-SRS method has the combination of one-electron and full potentials (for the wavefunction and energy corrections respectively). Since then, the R-SRS method provides stronger regularized induction and exchange induction energies than the RSEP-eN method. In addition, the RSEP-eN method gives non-regularized dispersion energy compared to the RSEP method (Figures \ref{fig:ch4-erf-disp-exdisp-plot}-\ref{fig:nh3clf-gau-disp-exdisp-plot}). 

The regularized first-order energy corrections show a systematic behavior in all tested systems as the separation parameter $\eta/\omega$ is varied. 
When this parameter approaches infinity, both RSEP and RSEP-eN recover the conventional SAPT0 values as expected. 
Interestingly, the electrostatic energies converge to the SAPT0 values from below for RSEP and from above for RSEP-eN for both range separations, that is, the RSEP method gives more attractive electrostatic energy compared to the conventional value. In all tested models, the long-ranged error function exhibits a faster energy convergence than the Gaussian function.
A desired feature of range separation is a relatively minor variance with the separation length parameter, indicating that the long-range Coulomb potential is responsible for nearly the entire noncovalent interaction.
From this perspective, considering all $\eta\in[2,\infty)$ bohr$^{-2}$ and $\omega\in[\sqrt{2},\infty)$ bohr$^{-1}$, the variance of both RSEP and RSEP-eN electrostatic energies is quite minor, with  full RSEP performing better for all systems except CH$_4\cdots$CH$_4$;  see the Supporting Information for more detail.

The first-order exchange energy for the RSEP-eN scheme converges from below with increasing $\eta/\omega$, in the opposite direction to electrostatic energy (thus, the range-separated sum $E^{(10)}_{\rm elst}+E^{(10)}_{\rm exch}$ exhibits some degree of error cancellation).
For the full range separation, the exchange energies for different $\eta/\omega$ are much closer to each other and to the full SAPT0 value. Nevertheless, the convergence is sometimes nonmonotonic, like for the $\text {NH}_{3} \cdots \text {ClF}$ system (the upper right panels in Figs.~\ref{fig:nh3clf-gau-rsep-plot} and \ref{fig:nh3clf-erf-rsep-plot}). 
%For example, for the $\text {NH}_{3} \cdots \text {ClF}$ system, the RSEP exchange energy at short range first increases away from the SAPT0 value, and then turns over at $\eta=3.0$ bohr$^{-2}$ and $\omega=\sqrt{3.0}$ bohr$^{-1}$, finally converging towards SAPT0 (see the upper right panel in Figs.~\ref{fig:nh3clf-gau-rsep-plot} and ~\ref{fig:nh3clf-erf-rsep-plot}).
%Note that, in this specific case, we included additionally the results for $\eta=15.0$ bohr$^{-2}$ and $\eta=20.0$ bohr$^{-2}$ to bridge the, still quite sizable, short-range discrepancy between conventional SAPT0 and the $\eta=10.0$ bohr$^{-2}$ data. 

The second-order induction and exchange-induction energies, as well as their sum $E^{(20)}_{\rm ind,resp}+E^{(20)}_{\rm exch-ind,resp}$, as functions of $\eta/\omega$ for all three systems are illustrated in the second row of Figs.~\ref{fig:h2o-gau-rsep-plot}--\ref{fig:nh3clf-erf-rsep-en-plot}.
The $\eta\to\infty$ $(\omega\to\infty)$ convergence is mostly monotonic -- the induction energy converges from above and the exchange-induction one converges from below, with a possible exception of the $\eta=1.0$ bohr$^{-2}$ and $\omega=1.0$ bohr$^{-1}$ data (the latter are highly inaccurate and might even exhibit a wrong sign).
These directions of convergence imply that the sum $E^{(20)}_{\rm ind,resp}+E^{(20)}_{\rm exch-ind,resp}$ exhibits some error cancellation; indeed, the range-separated sums are not only much smaller in magnitude than the separate $E^{(20)}_{\rm ind,resp}$ and $E^{(20)}_{\rm exch-ind,resp}$ terms (as commonly observed in standard SAPT), but their range of variability is usually also much smaller.

For all systems and both range separation schemes, the RSEP induction and exchange-induction terms are highly consistent between different values of $\eta/\omega$, exhibiting much less variance than RSEP-eN. 
In contrast, RSEP-eN leads to a smaller variance for the sum of these two terms for the methane dimer and the ionic bonding system H$_2$O$\cdots$F$^-$ in both range separation functions.
For the other systems, the sum exhibits less variance with the full RSEP range separation: by a minimal amount for H$_2$O$\cdots$H$_2$O and by a significant factor for NH$_3\cdots$ClF.
Overall, the RSEP-eN variant could likely perform better for estimating the charge-transfer energy \cite{Misquitta:13} as a larger reduction in magnitude of the mutually cancelling induction and exchange-induction terms is afforded. 
For the purpose of approximating the interaction potential by its long-range term, RSEP appears to offer superior accuracy, in particular for the potentially most difficult NH$_3\cdots$ClF complex. 

We now move on to the range-separated second-order dispersion and exchange-dispersion energies displayed in the middle right panel of Figs.~\ref{fig:h2o-gau-rsep-plot}--\ref{fig:nh3clf-erf-rsep-en-plot}.
As expected, all RSEP-eN dispersion terms are virtually identical to the conventional SAPT0 dispersion, and the corresponding RSEP-eN exchange-dispersion data exhibit very little variance with respect to $\eta/\omega$.
This behavior is due to the matrix elements involved in the computation of dispersion corrections not depending on the one-electron integrals.
In contrast, the RSEP corrections do exhibit some variance with $\eta/\omega$, monotonically converging to the conventional SAPT0 data from above for $E^{(20)}_{\rm disp}$ and from below for $E^{(20)}_{\rm exch-disp}$.
The cancellation of range-separation effects in the sum $E^{(20)}_{\rm disp}+E^{(20)}_{\rm exch-disp}$ is reasonable but not quite perfect, and the RSEP sum exhibits some variation with the separation parameter, especially for the Gaussian regularization. In all models, the error function provides a better consistency on the total dispersion energy.
Nevertheless, in the practically important range $\eta=3.0$ bohr$^{-2}$ and above, RSEP can produce reasonably accurate (dispersion plus exchange-dispersion) energies at the van der Waals minima and beyond.

The total range-separated SAPT0-level interaction energies for our four example systems are presented at the bottom center of Figs.~\ref{fig:h2o-gau-rsep-plot}--\ref{fig:nh3clf-erf-rsep-en-plot} and in the Supporting Information.
As in Eq.~(\ref{eq:sapt0}), the total interaction energy includes the delta Hartree-Fock term that is taken here from the calculation without range separation.
For all systems and range separation schemes, the differences in electrostatic energy and in first-order exchange energy, partially balancing each other, have the biggest contribution to the differences between range-separated and conventional SAPT0.
For the $\text {NH}_3 \cdots \text {ClF}$ system at the shortest $R$, as the monomers get a bit closer to each other than the equilibrium distance, most range separated SAPT variants erroneously give a repulsive total interaction energy.
In this case, the full error function separation gives an attractive total interaction energy as long as $\omega > \sqrt{3.0}$ bohr$^{-1}$. 
In comparison, in the full Gaussian range separation as well as in the RSEP-eN method and either scheme, the attractive total energy is not observed even when $\eta = 10.0$ bohr$^{-2}$ (see Figures \ref{fig:nh3clf-gau-rsep-plot} and \ref{fig:nh3clf-gau-rsep-en-plot}).
Overall, the full RSEP range separation using the error function provides the best agreement with conventional SAPT0 for a wide range of $\omega$ values.   
The Gaussian range separation with parameter $\eta$ typically shows somewhat larger total energy discrepancies than the error function one with $\omega=\sqrt{\eta}$, but still exhibits acceptable accuracy in a wide range of $\eta$.
The RSEP-eN scheme in general gives larger total energy errors than the RSEP one, indicating that a more balanced description of the attractive and repulsive Coulomb interactions is more accurate than a milder but unbalanced approximation.

\begin{figure}[!htb]
\begin{minipage}{\textwidth}
\centering
\includegraphics[width=1\textwidth]{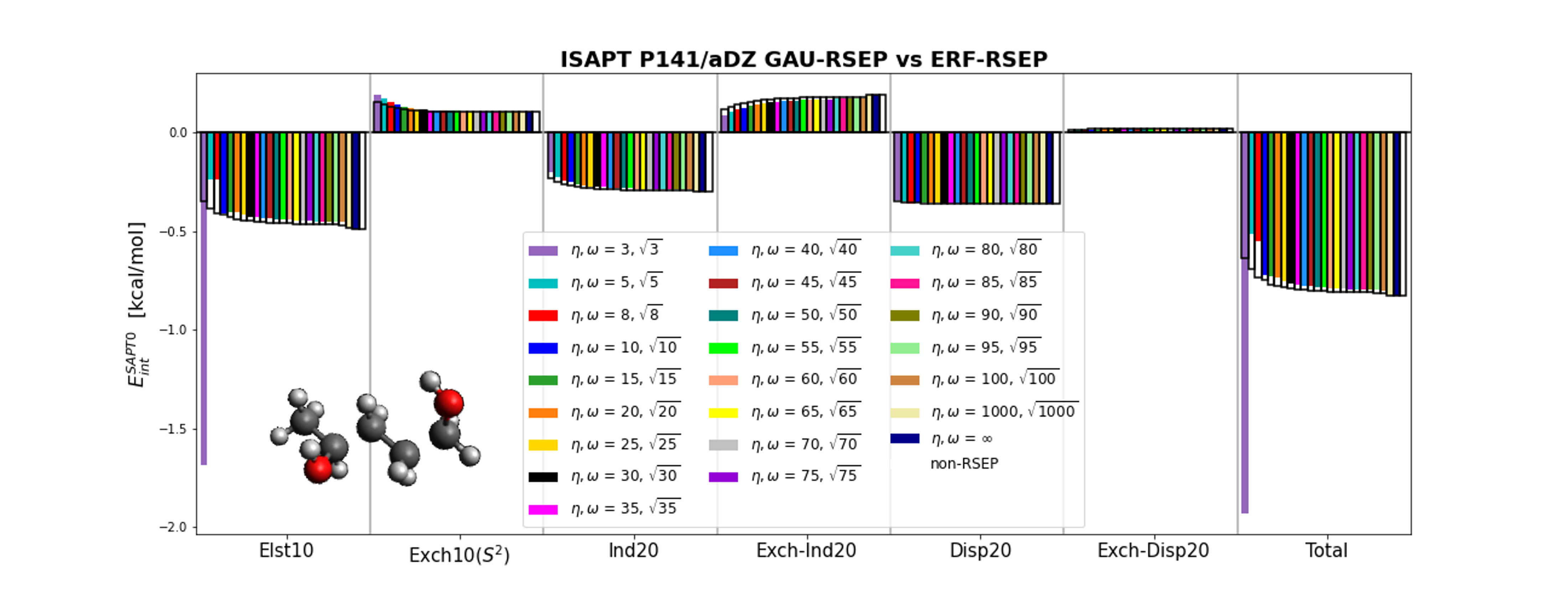}
\end{minipage}
\caption{The components of the range-separated ISAPT/SIAO1 interaction energy of the 1,4-pentanediol molecule (fragmented as pictured) computed with different values of $\eta$ (bohr$^{-2}$, colored bars) and $\omega$ (bohr$^{-1}$, empty bars) in the aDZ basis set. The energy results are compared to those of the traditional ISAPT/SIAO1 method.}
\label{fig:P141-adz-gau-erf-memdf-plot}
\end{figure}

\begin{figure}[!htb]
\begin{minipage}{.48\textwidth}
\centering
\includegraphics[width=1\textwidth]{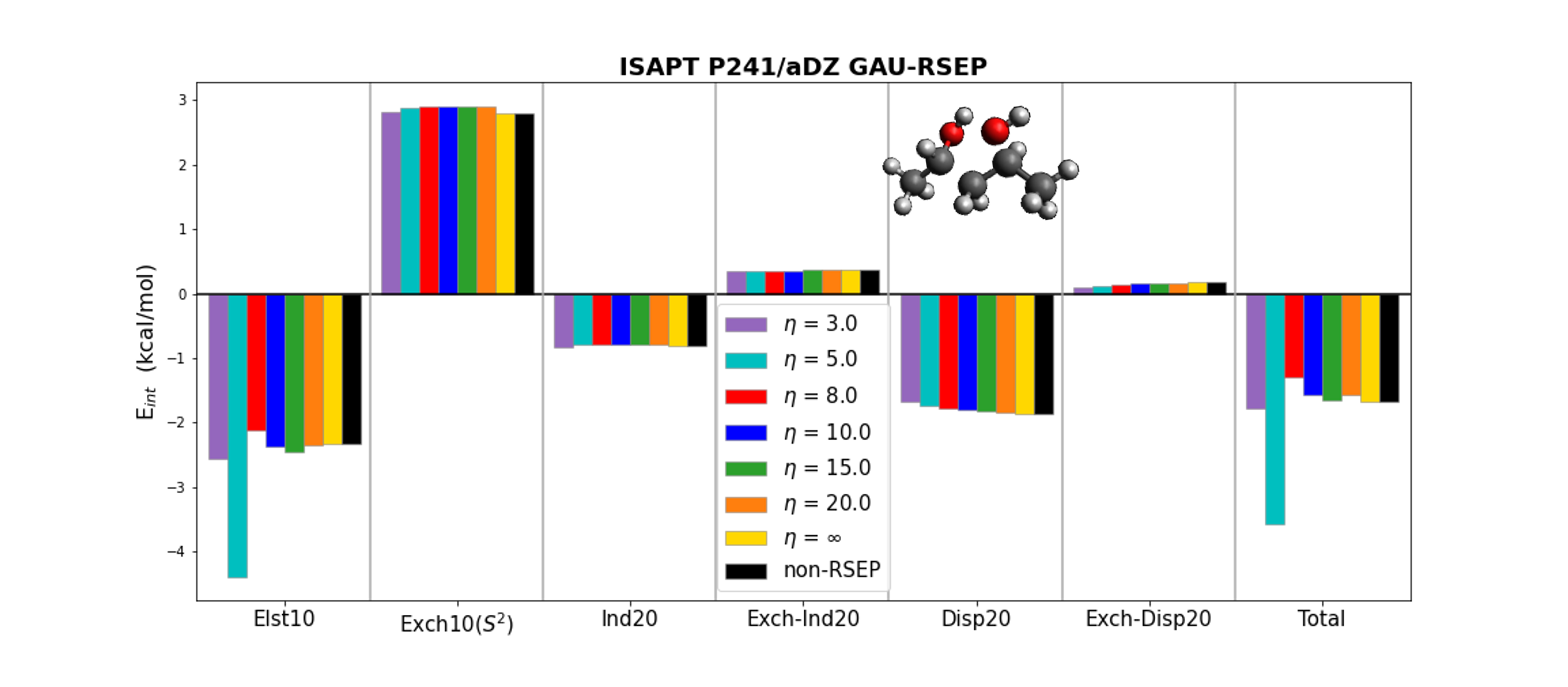}
\end{minipage}
\begin{minipage}{.48\textwidth}
\centering
\includegraphics[width=1\textwidth]{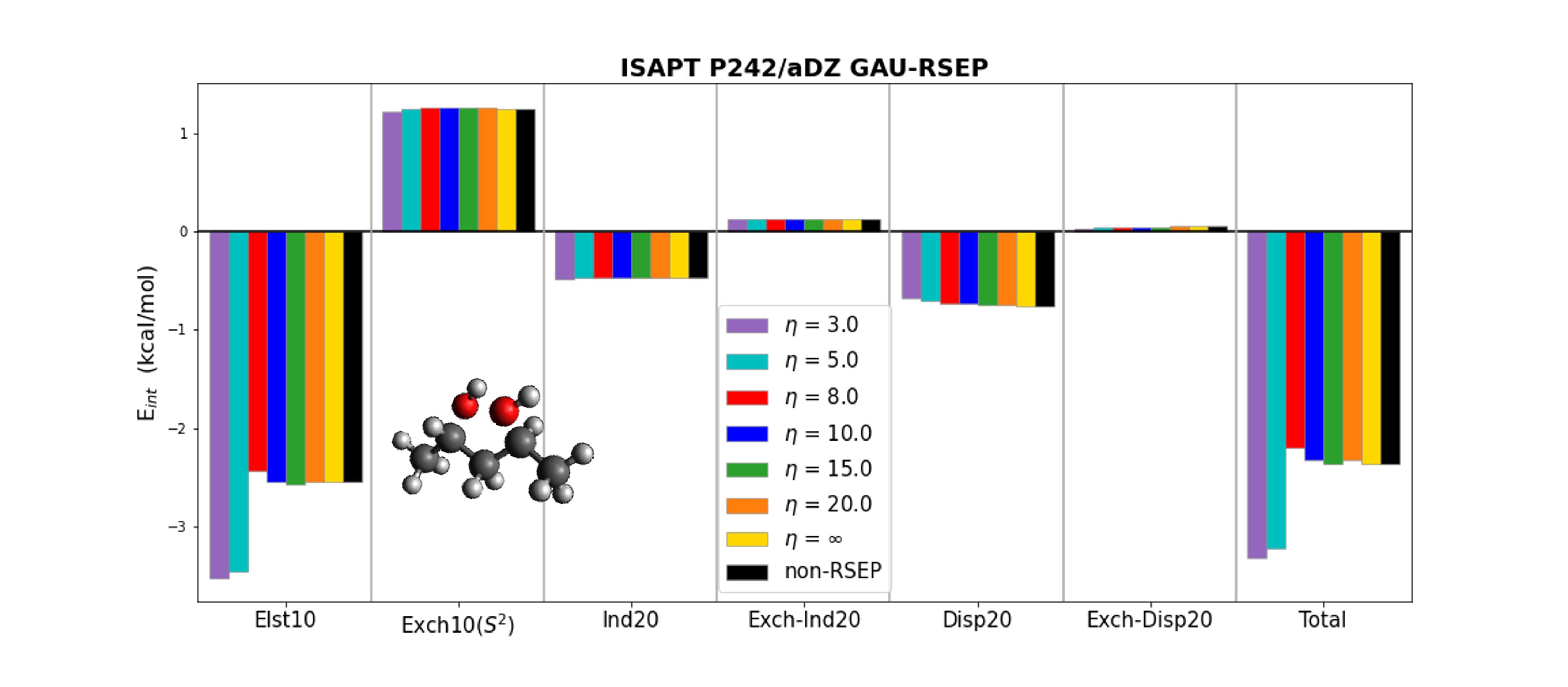}
\end{minipage}
\begin{minipage}{.48\textwidth}
\centering
\includegraphics[width=1\textwidth]{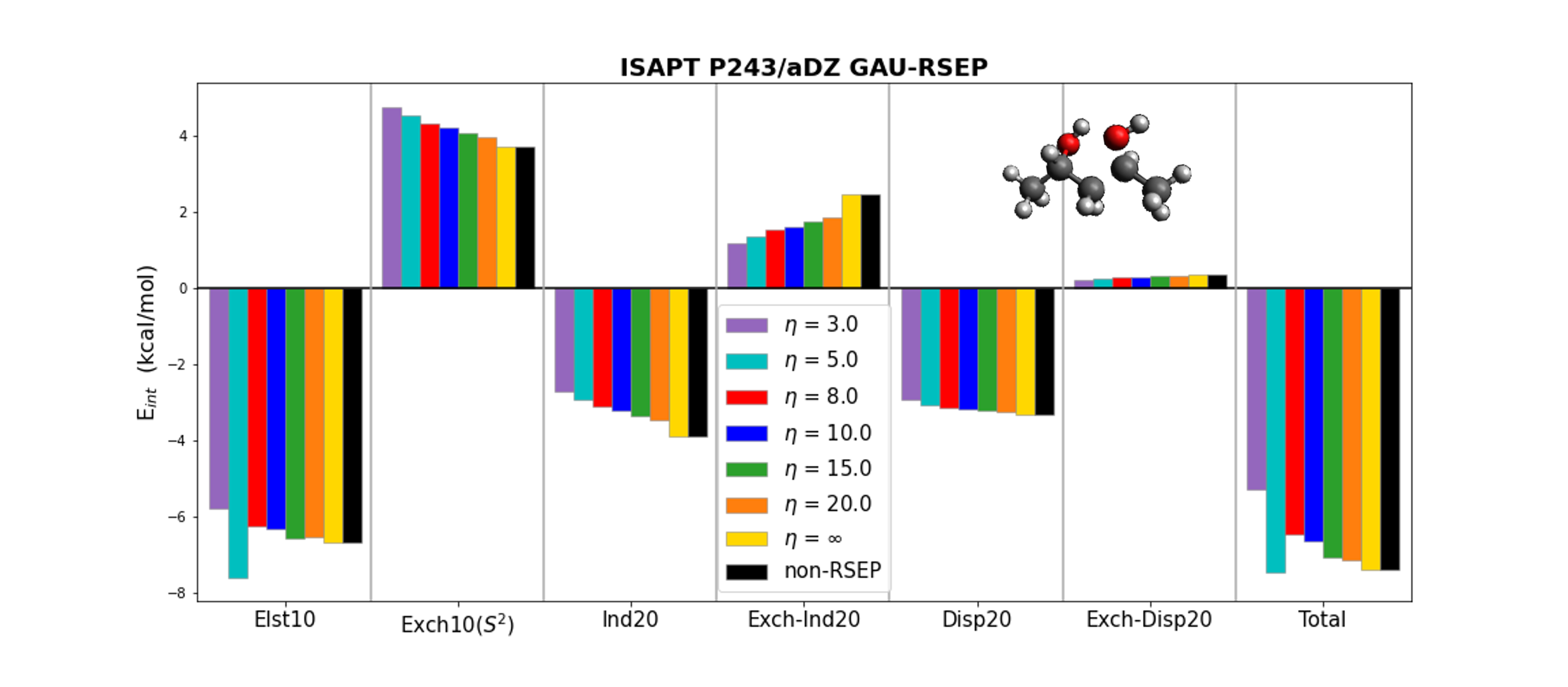}
\end{minipage}
\begin{minipage}{.48\textwidth}
\centering
\includegraphics[width=1\textwidth]{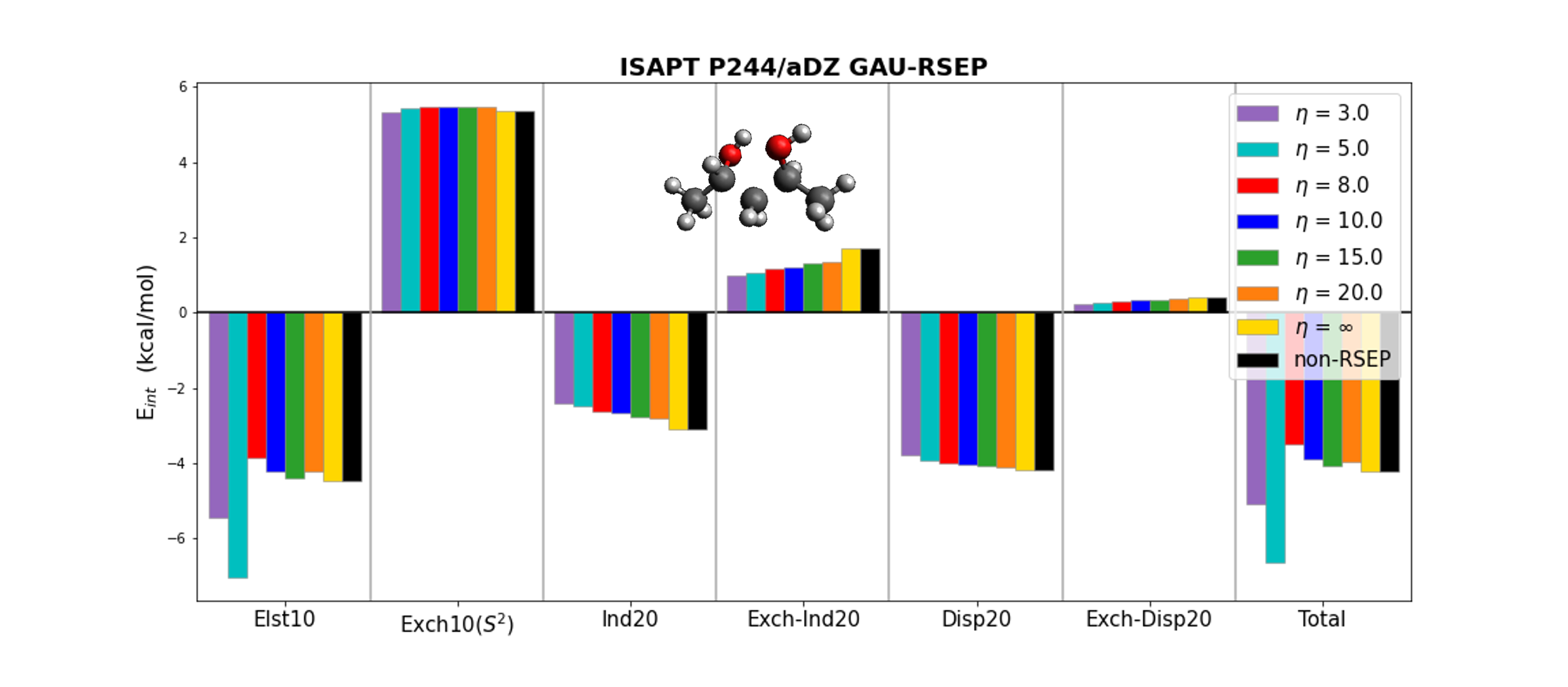}
\end{minipage}
\caption{The components of the ISAPT/SIAO1 (RSEP/gau) interaction energy for 4 fragmentation models (pictured) of the 2,4-pentanediol molecule computed with different values of $\eta$ (bohr$^{-2}$) in the aDZ basis set. The energy results are compared to those of the traditional ISAPT/SIAO1 method.}
\label{fig:24-c5h1202-gau-memdf-plots}
\end{figure}

\begin{figure}[!htb]
\begin{minipage}{.48\textwidth}
\centering
\includegraphics[width=1\textwidth]{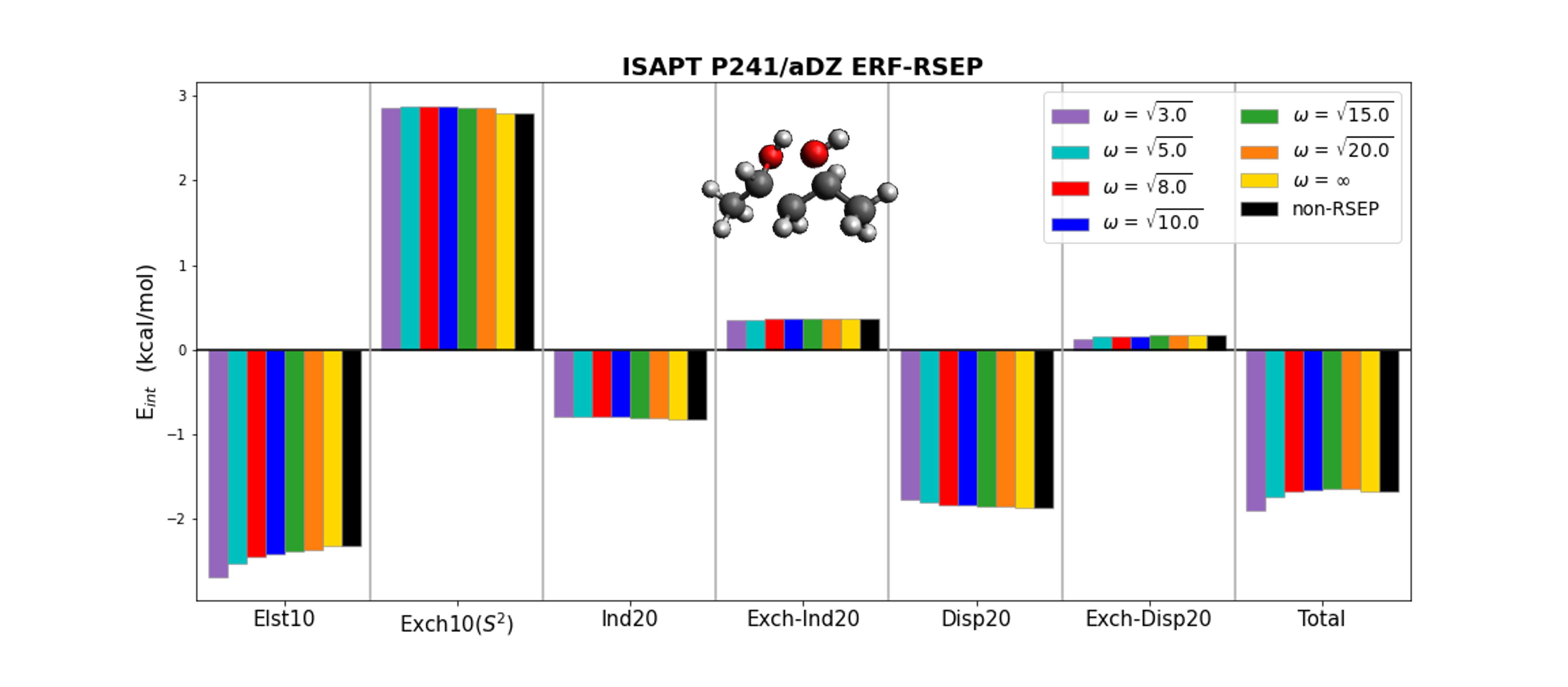}
\end{minipage}
\begin{minipage}{.48\textwidth}
\centering
\includegraphics[width=1\textwidth]{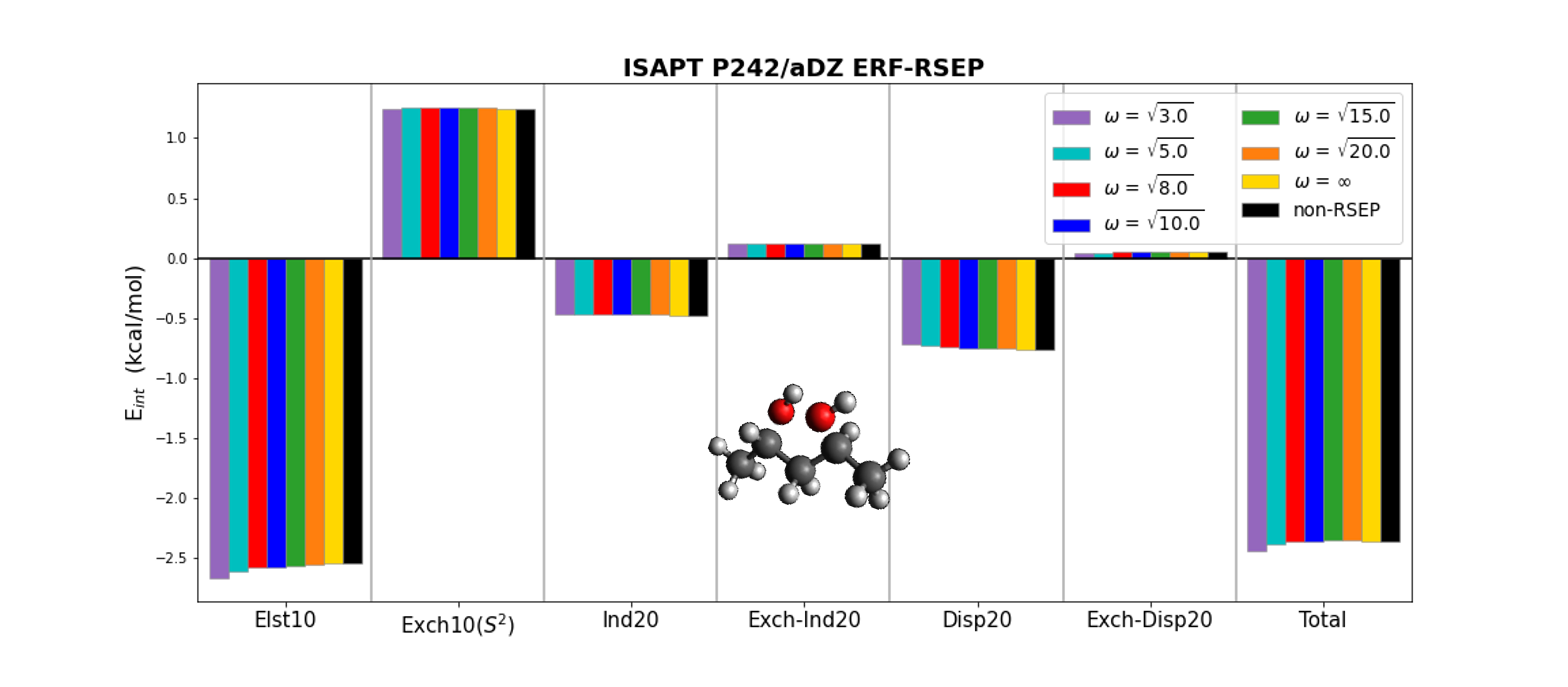}
\end{minipage}
\begin{minipage}{.48\textwidth}
\centering
\includegraphics[width=1\textwidth]{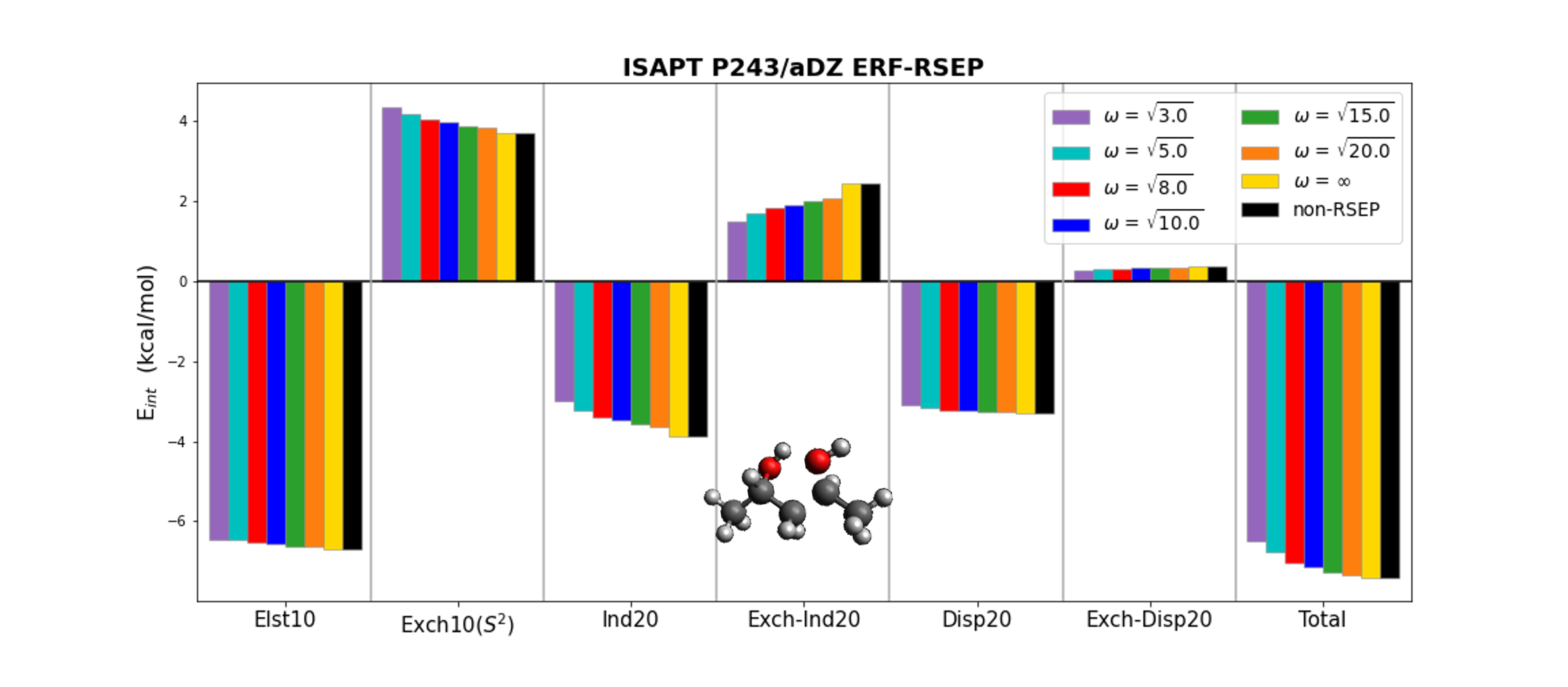}
\end{minipage}
\begin{minipage}{.48\textwidth}
\centering
\includegraphics[width=1\textwidth]{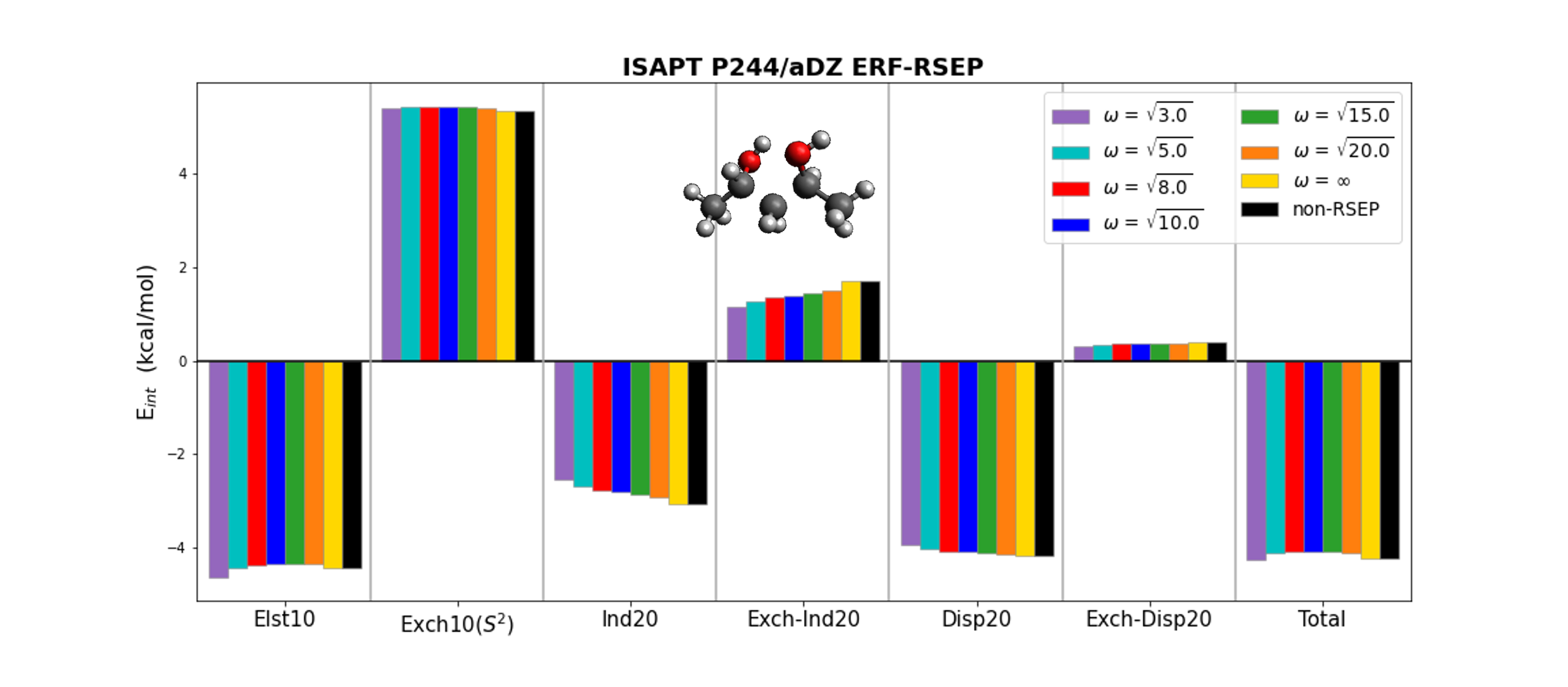}
\end{minipage}
\caption{The components of the ISAPT/SIAO1 (RSEP/erf) interaction energy for 4 fragmentation models (pictured) of the 2,4-pentanediol molecule computed with different values of $\omega$ (bohr$^{-1}$) in the aDZ basis set. The energy results are compared to those of the traditional ISAPT/SIAO1 method.}
\label{fig:24-c5h1202-erf-memdf-plots}
\end{figure}

\begin{figure}[!htb]
\begin{minipage}{.48\textwidth}
\centering
\includegraphics[width=1\textwidth]{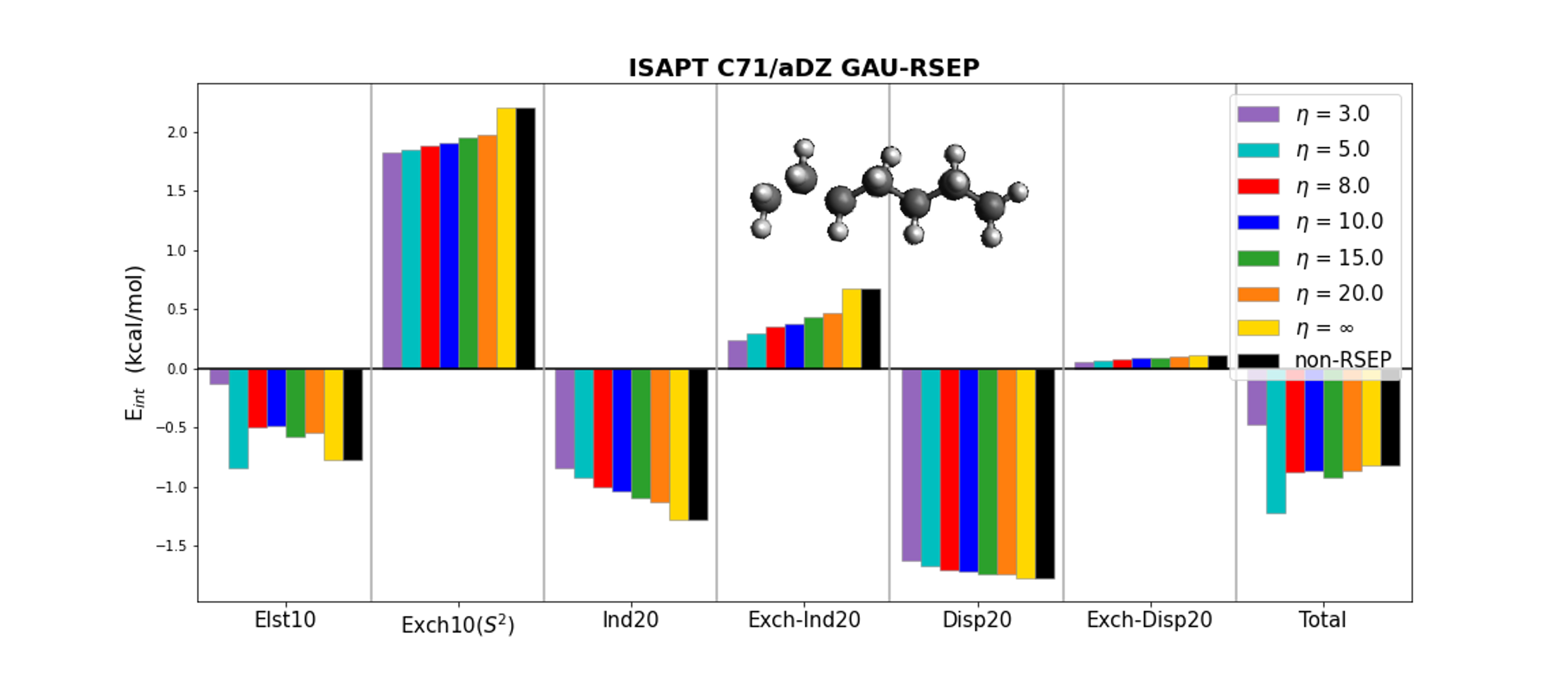}
\end{minipage}
\vspace{-.22in}
\begin{minipage}{.48\textwidth}
\centering
\includegraphics[width=1\textwidth]{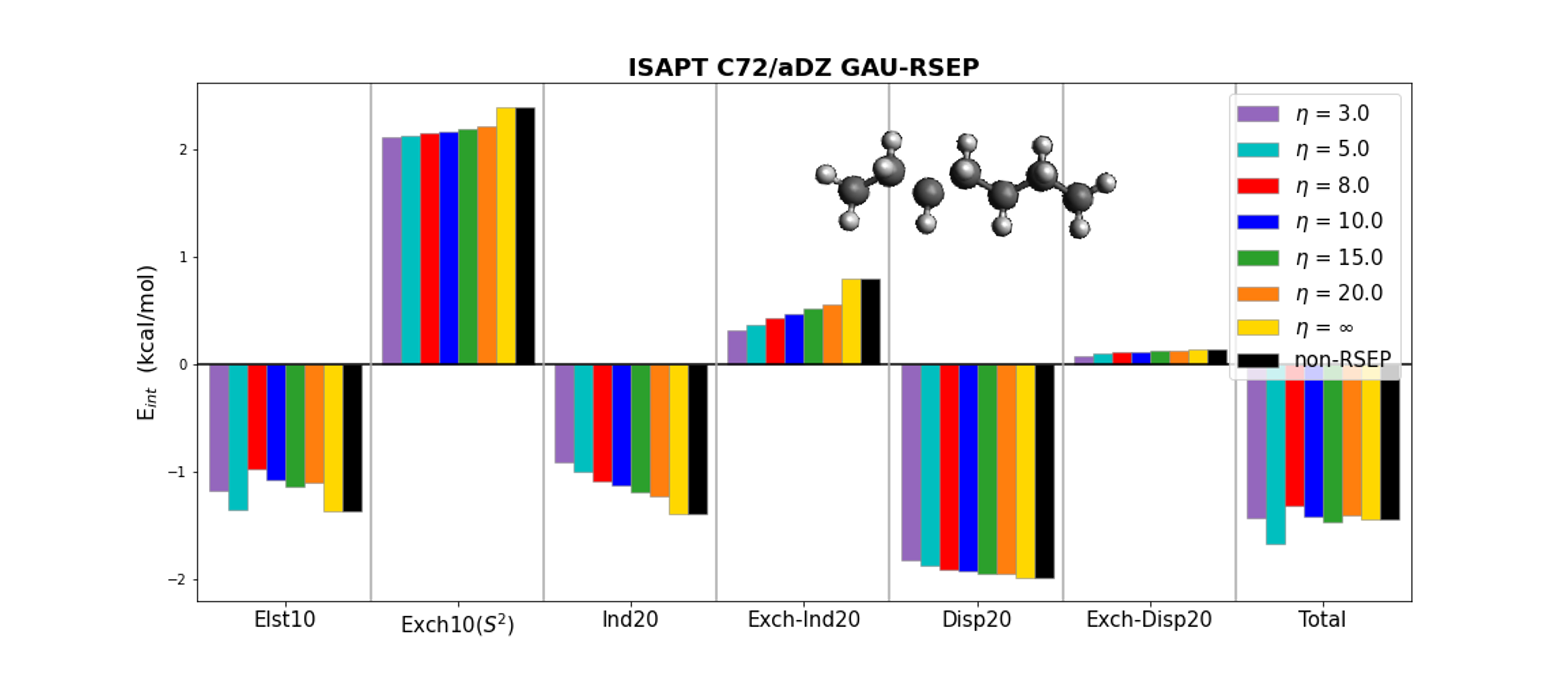}
\end{minipage}
\vspace{-.22in}
\begin{minipage}{.48\textwidth}
\centering
\includegraphics[width=1\textwidth]{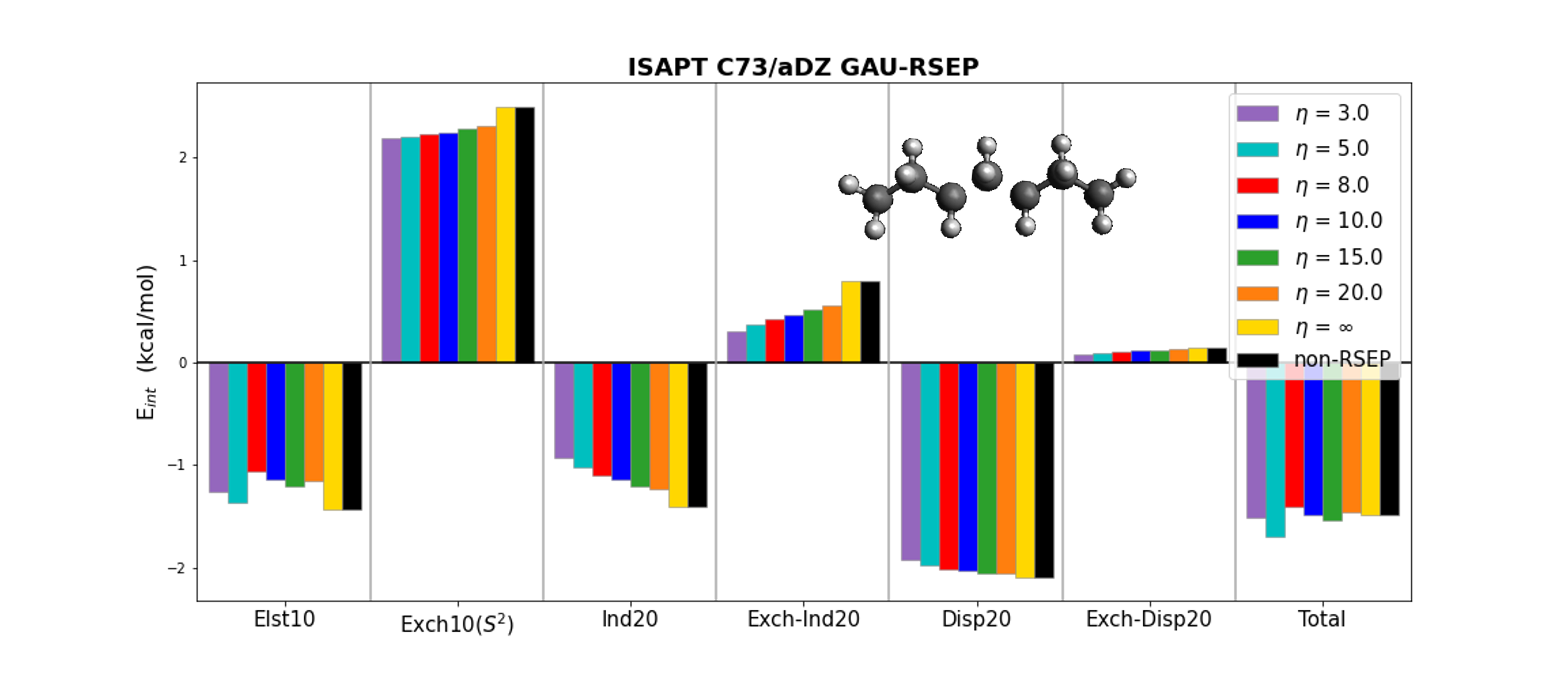}
\end{minipage}
\vspace{-.01in}
\begin{minipage}{.48\textwidth}
\centering
\includegraphics[width=1\textwidth]{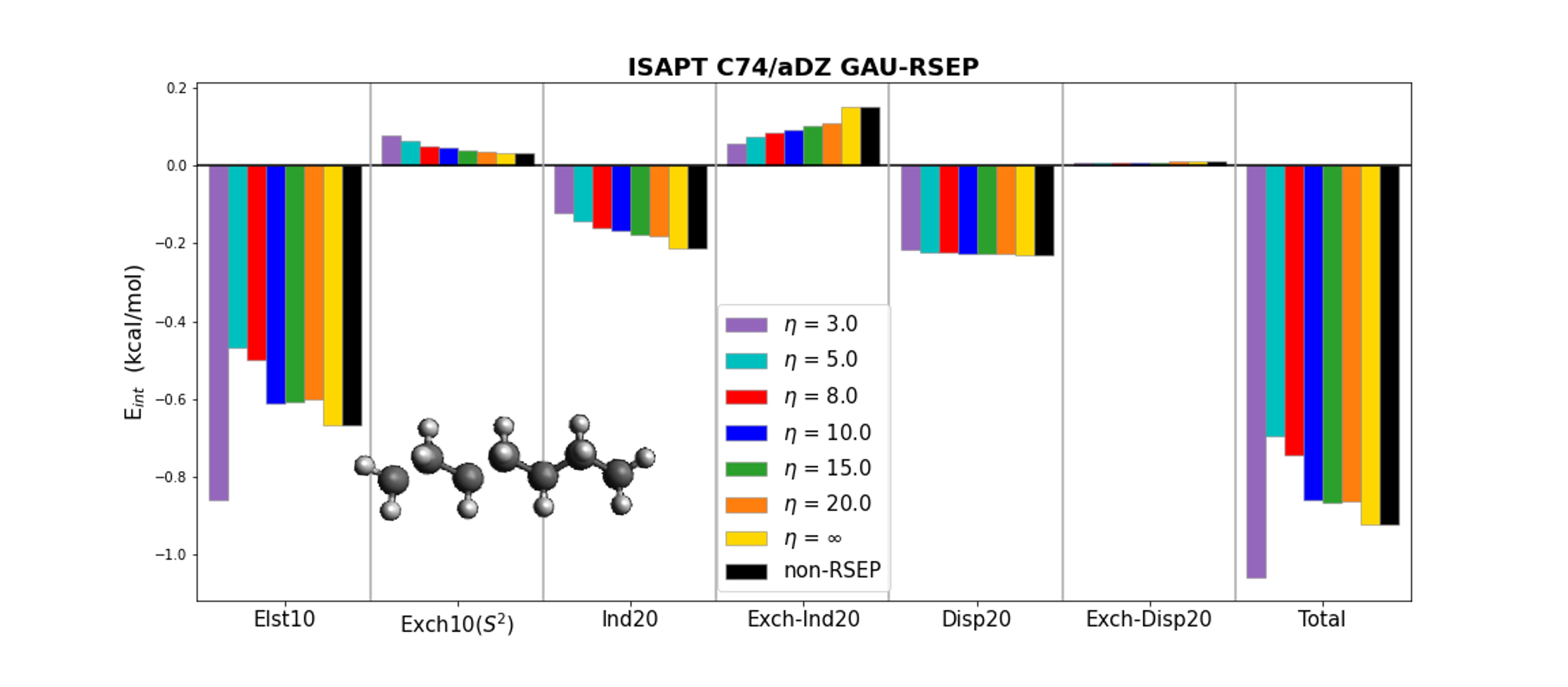}
\end{minipage}
\vspace{-.01in}
\begin{minipage}{.48\textwidth}
\centering
\includegraphics[width=1\textwidth]{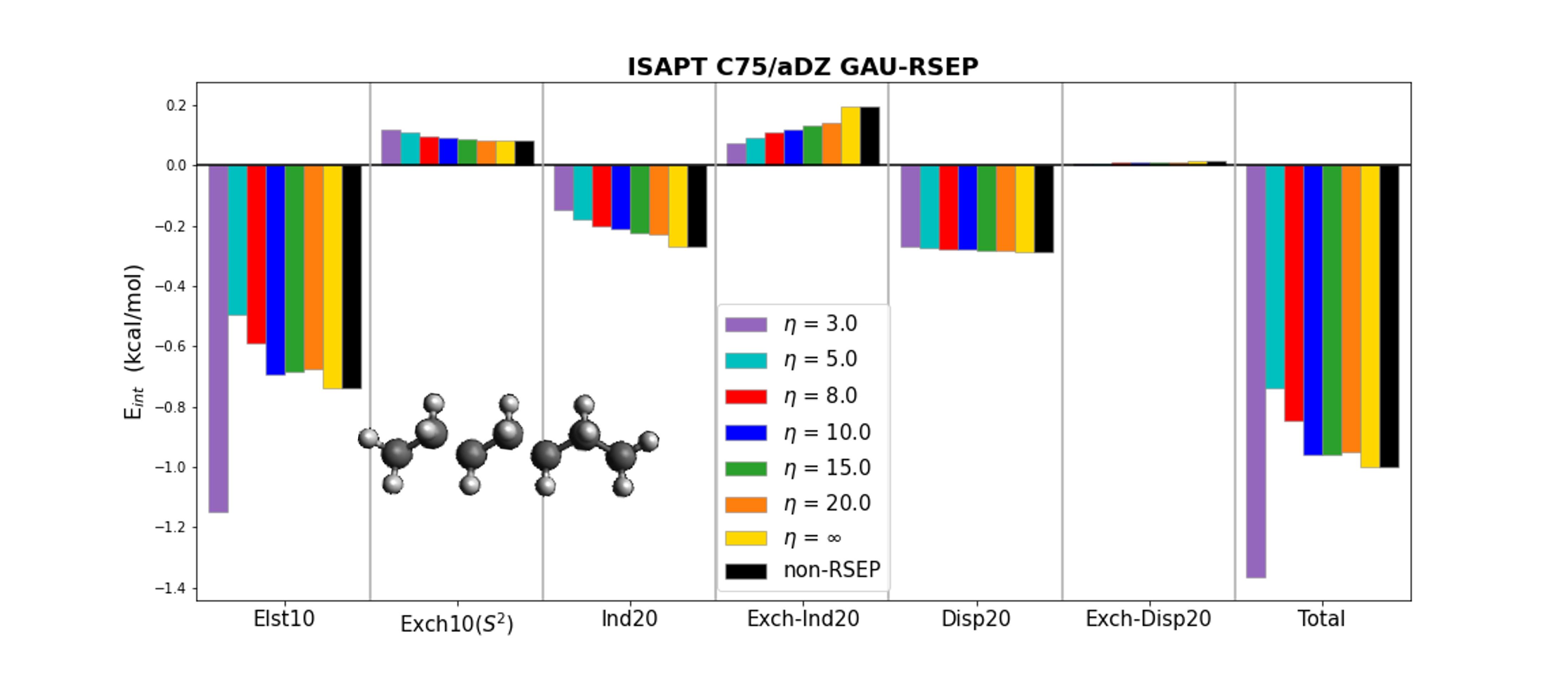}
\end{minipage}
\vspace{-.21in}
\begin{minipage}{.48\textwidth}
\centering
\includegraphics[width=1\textwidth]{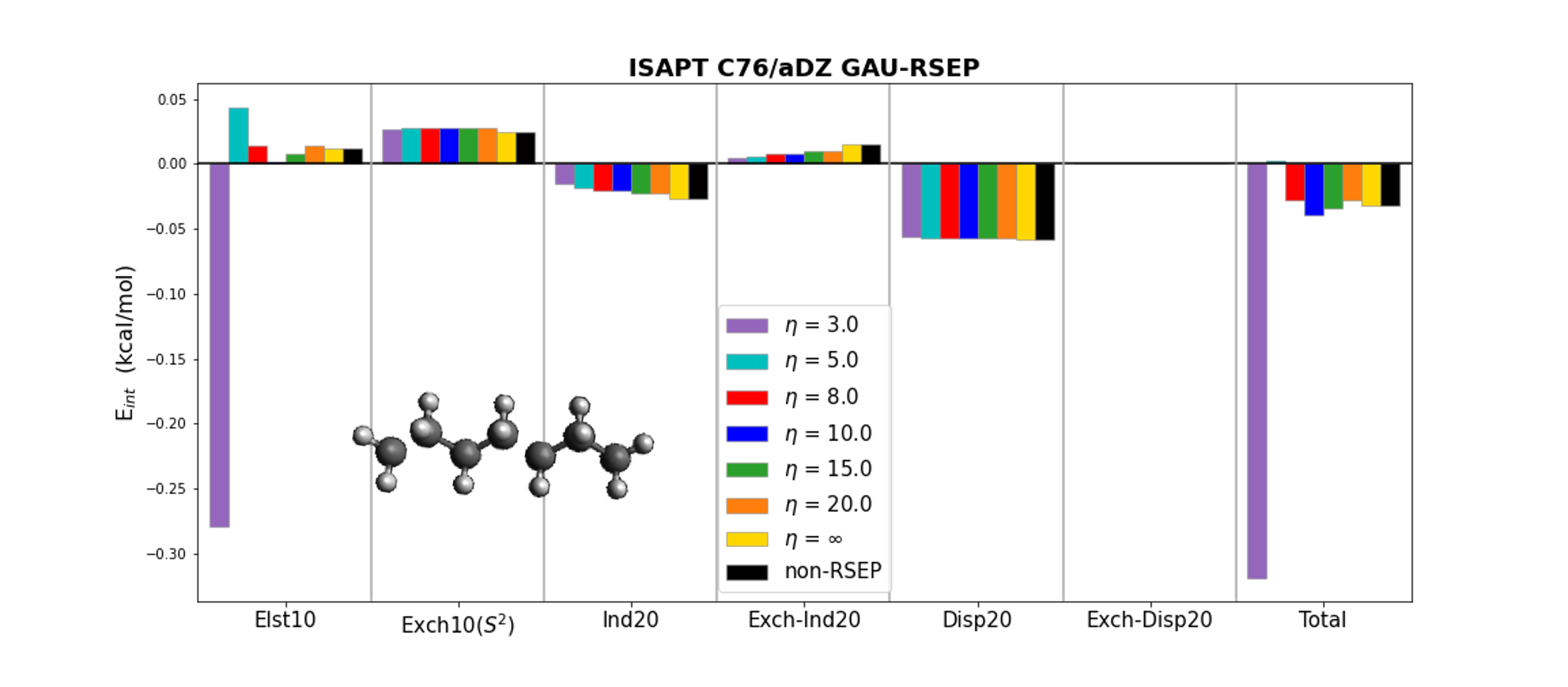}
\end{minipage}
\vspace{-.21in}
\begin{minipage}{.48\textwidth}
\centering
\includegraphics[width=1\textwidth]{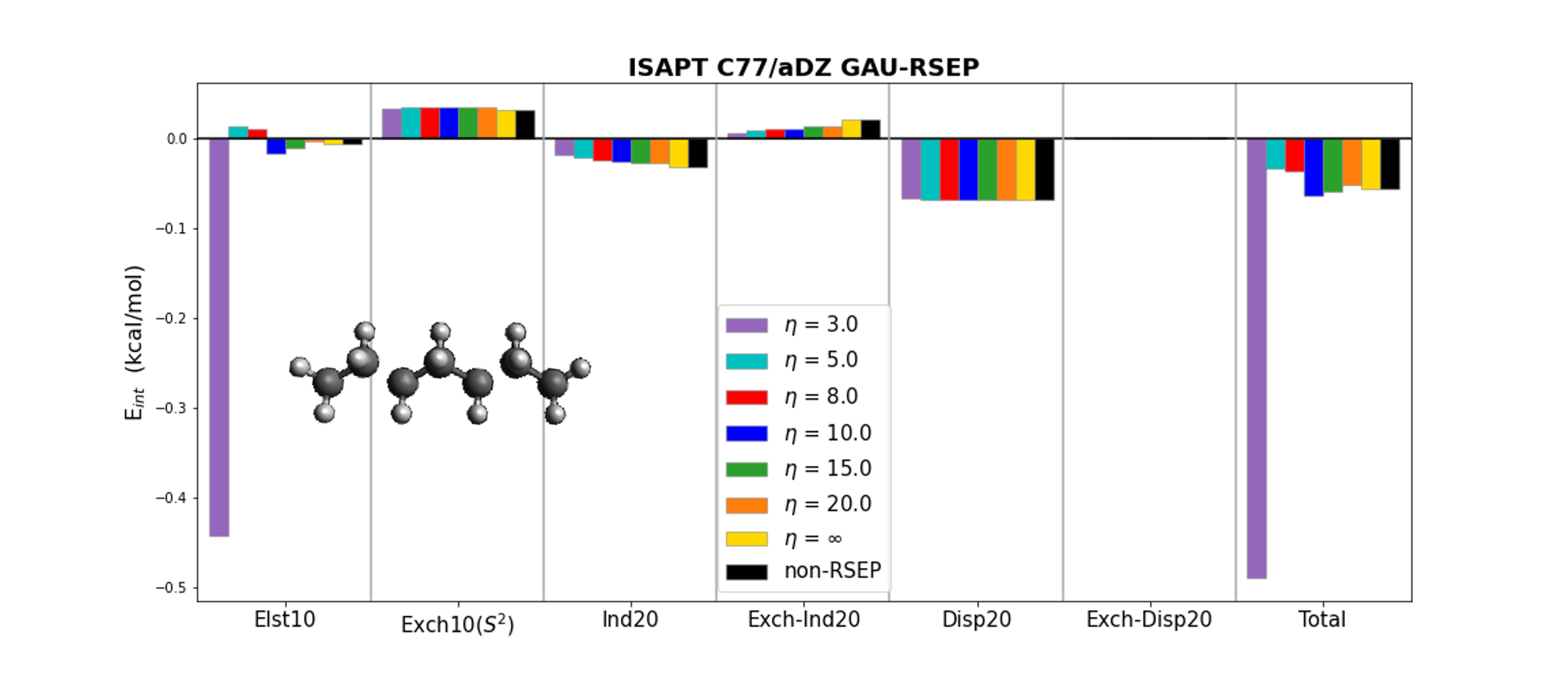}
\end{minipage}
\vspace{-.02in}
\begin{minipage}{.48\textwidth}
\centering
\includegraphics[width=1\textwidth]{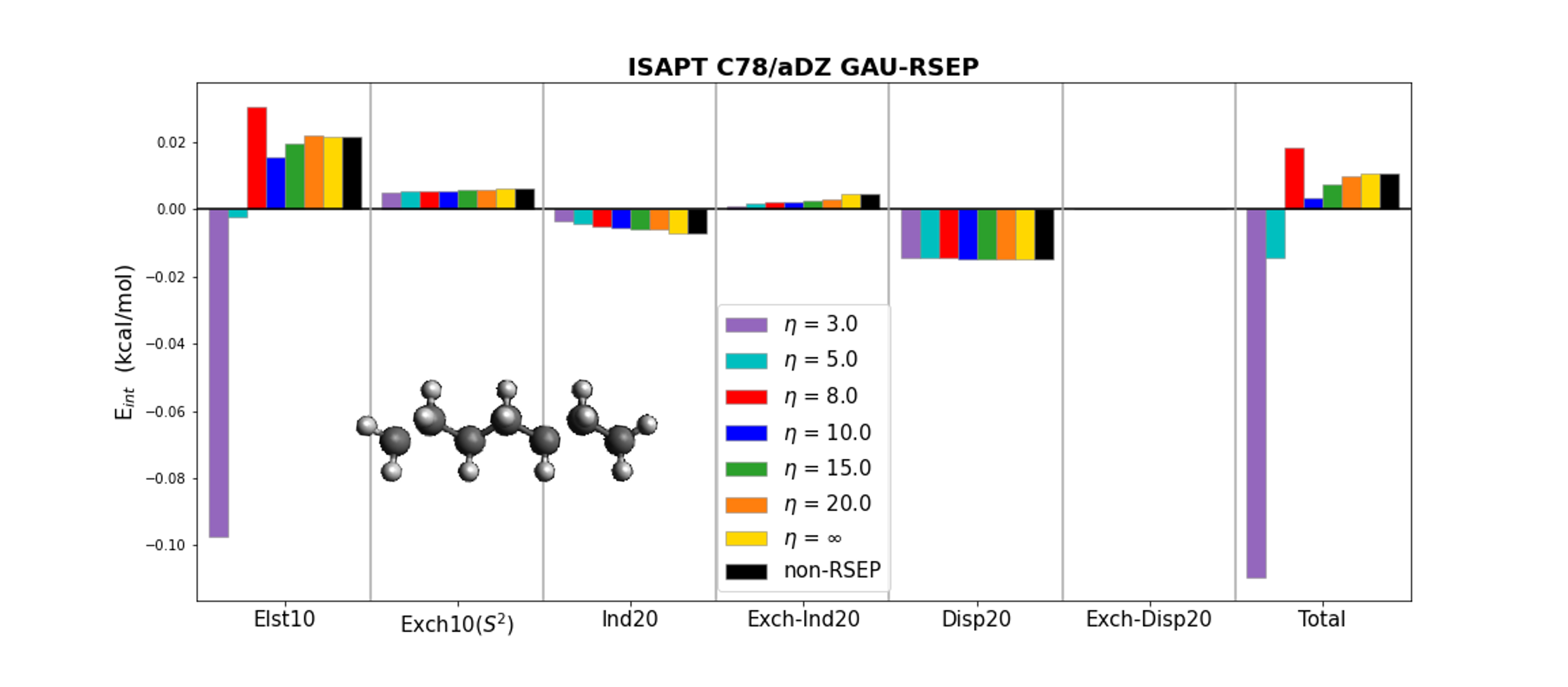}
\end{minipage}
\vspace{-.02in}
\begin{minipage}{.48\textwidth}
\centering
\includegraphics[width=1\textwidth]{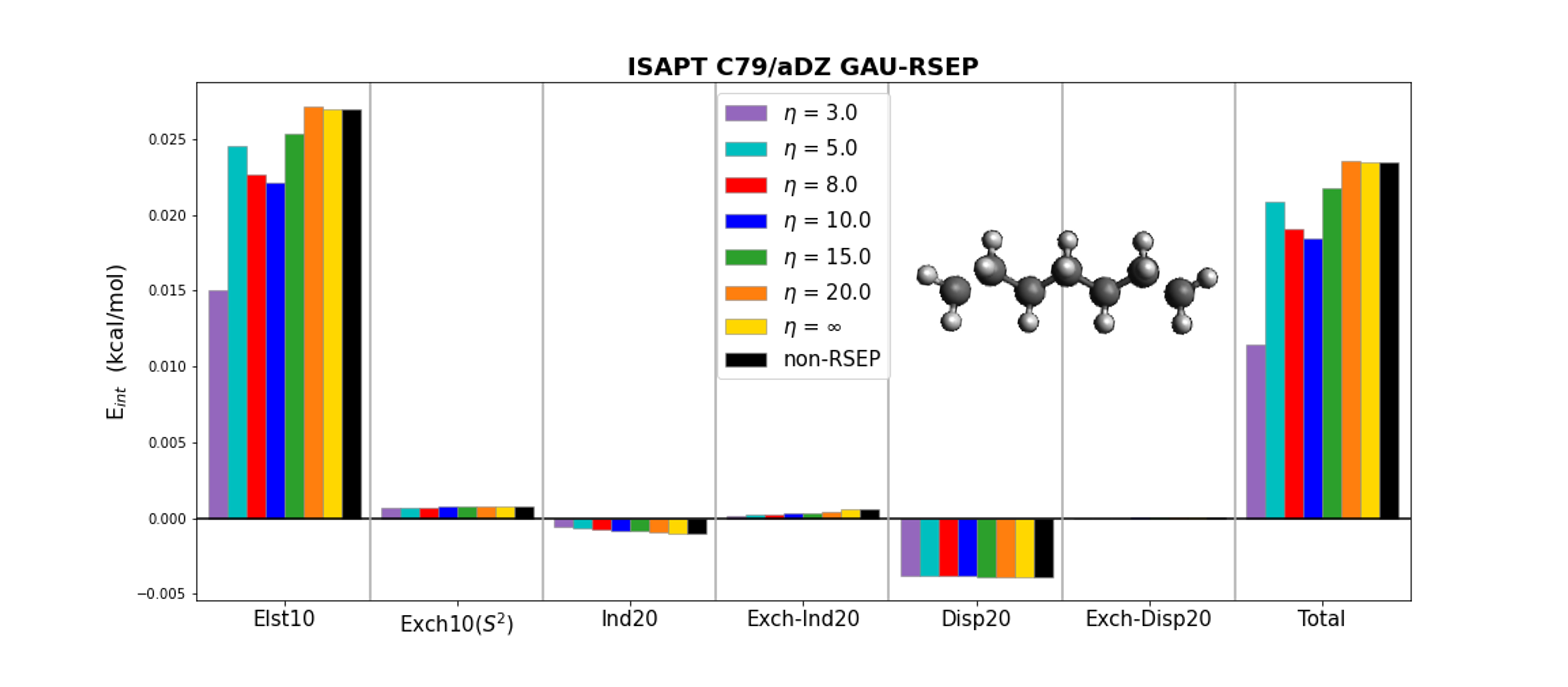}
\end{minipage}
\caption{The components of the ISAPT/SIAO1 (RSEP/gau) interaction energy for 9 fragmentation models (pictured) of the n-heptane molecule computed with different values of $\eta$ (bohr$^{-2}$) in the aDZ basis set. The energy results are compared to those of the traditional ISAPT/SIAO1 method.}
\label{fig:c7h16-gau-memdf-plots}
\end{figure}

\begin{figure}[!htb]
\begin{minipage}{.48\textwidth}
\centering
\includegraphics[width=1\textwidth]{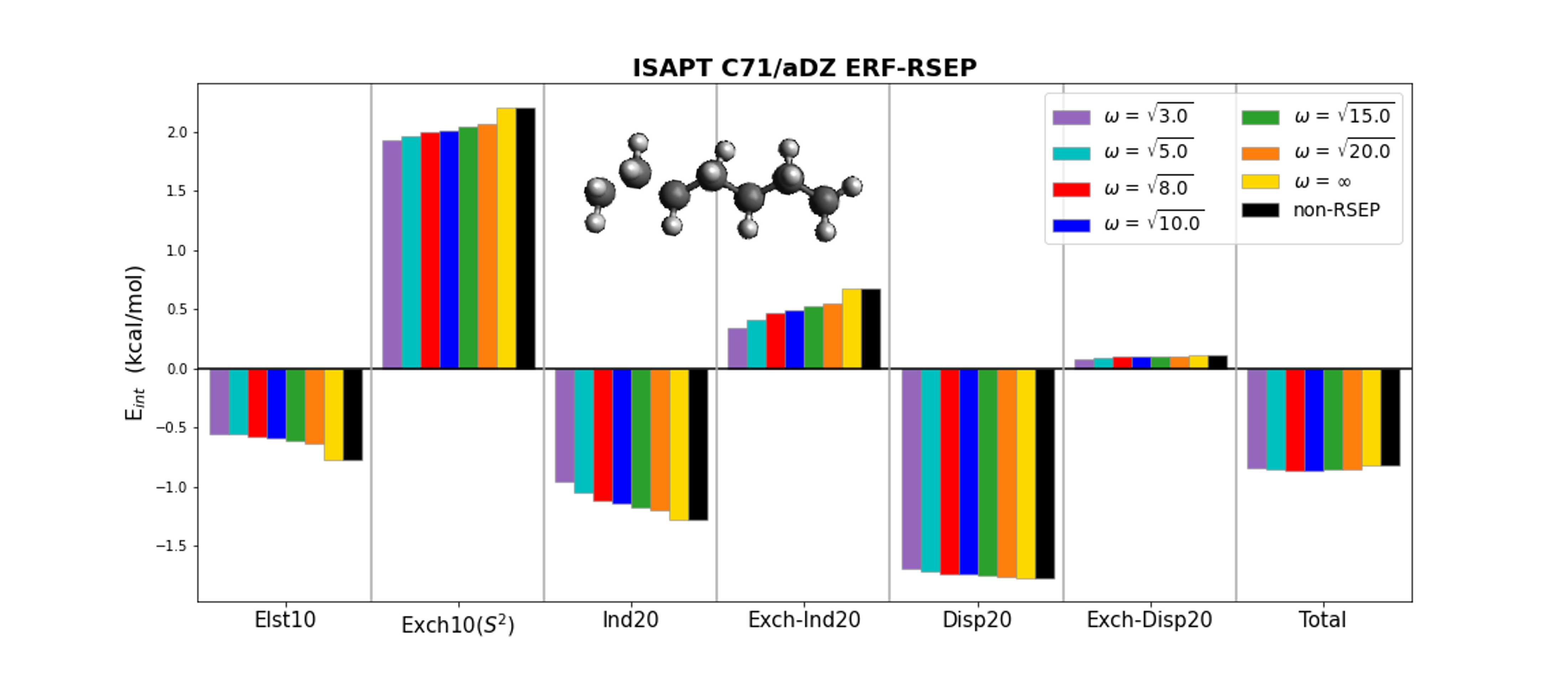}
\end{minipage}
\vspace{-.22in}
\begin{minipage}{.48\textwidth}
\centering
\includegraphics[width=1\textwidth]{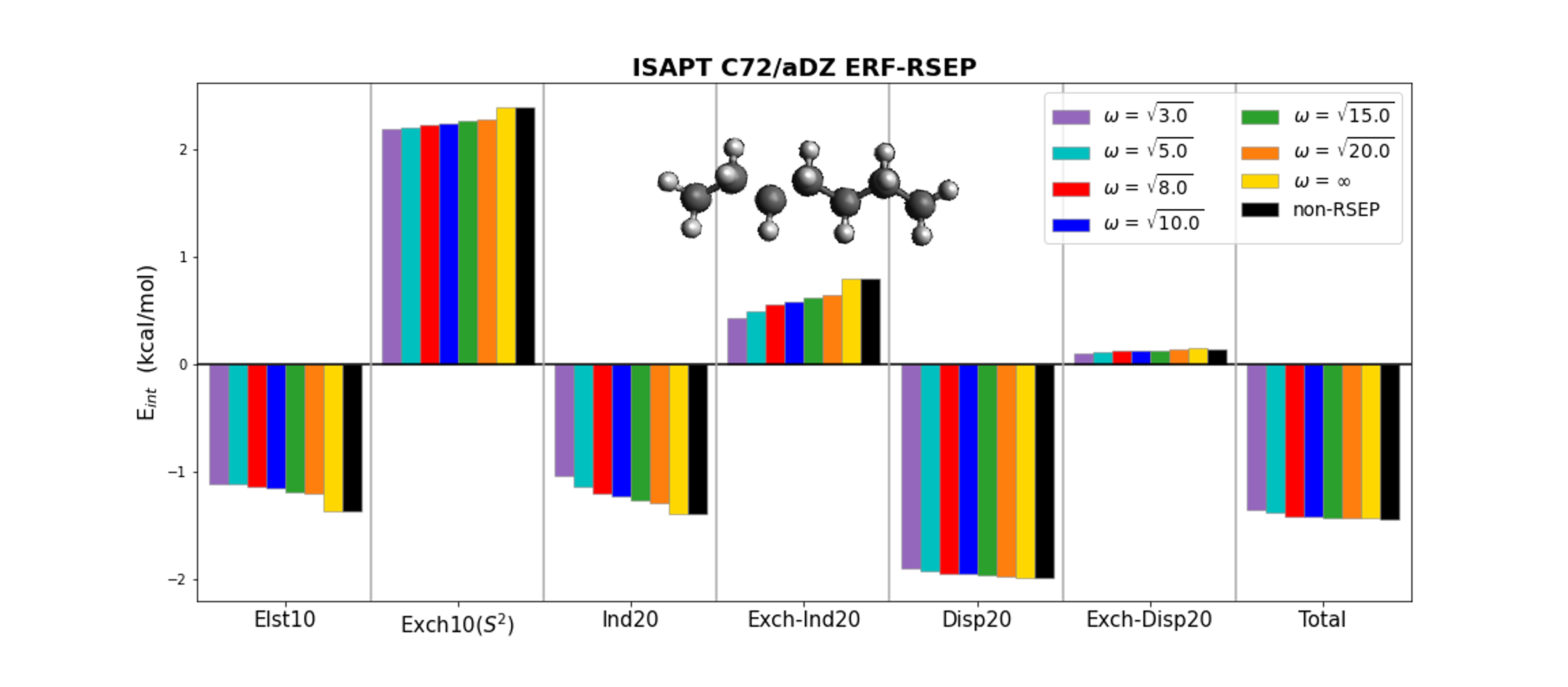}
\end{minipage}
\vspace{-.22in}
\begin{minipage}{.48\textwidth}
\centering
\includegraphics[width=1\textwidth]{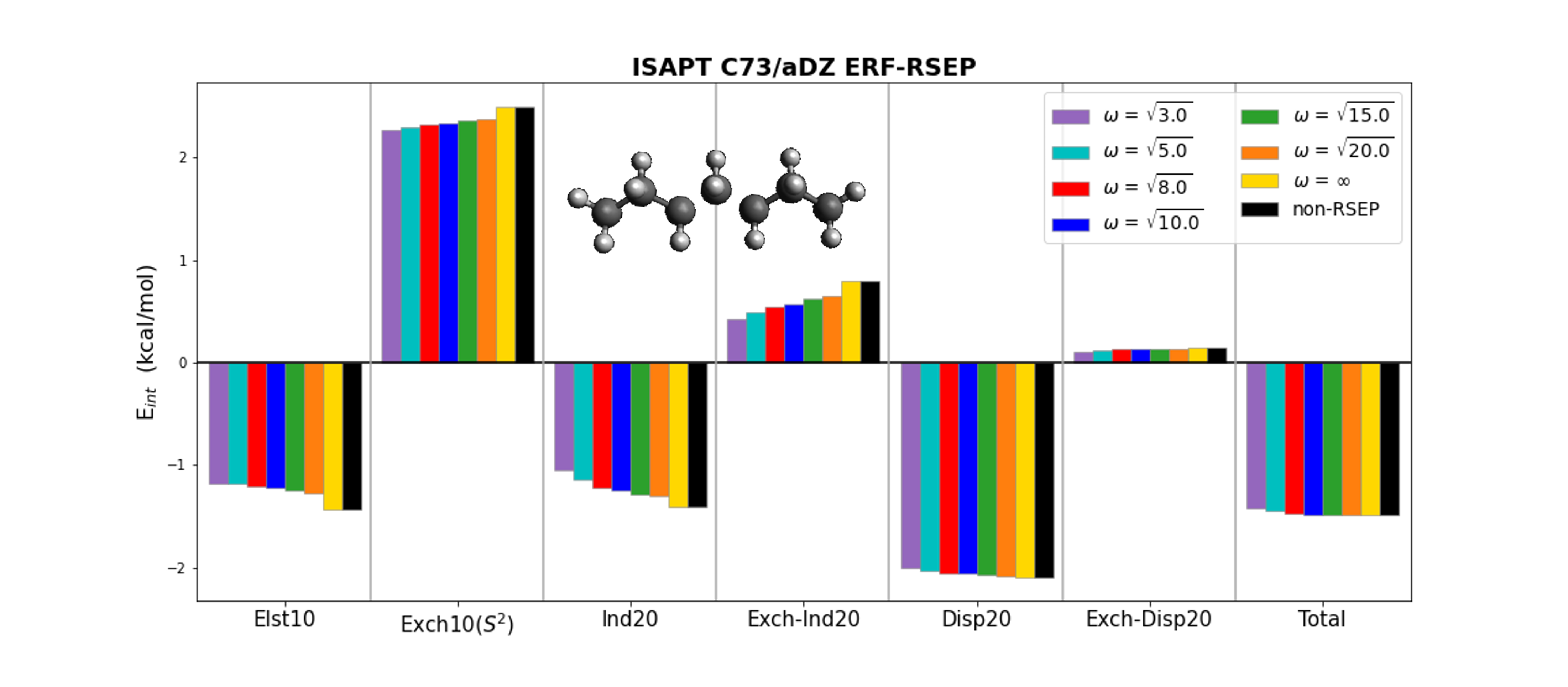}
\end{minipage}
\vspace{-.01in}
\begin{minipage}{.48\textwidth}
\centering
\includegraphics[width=1\textwidth]{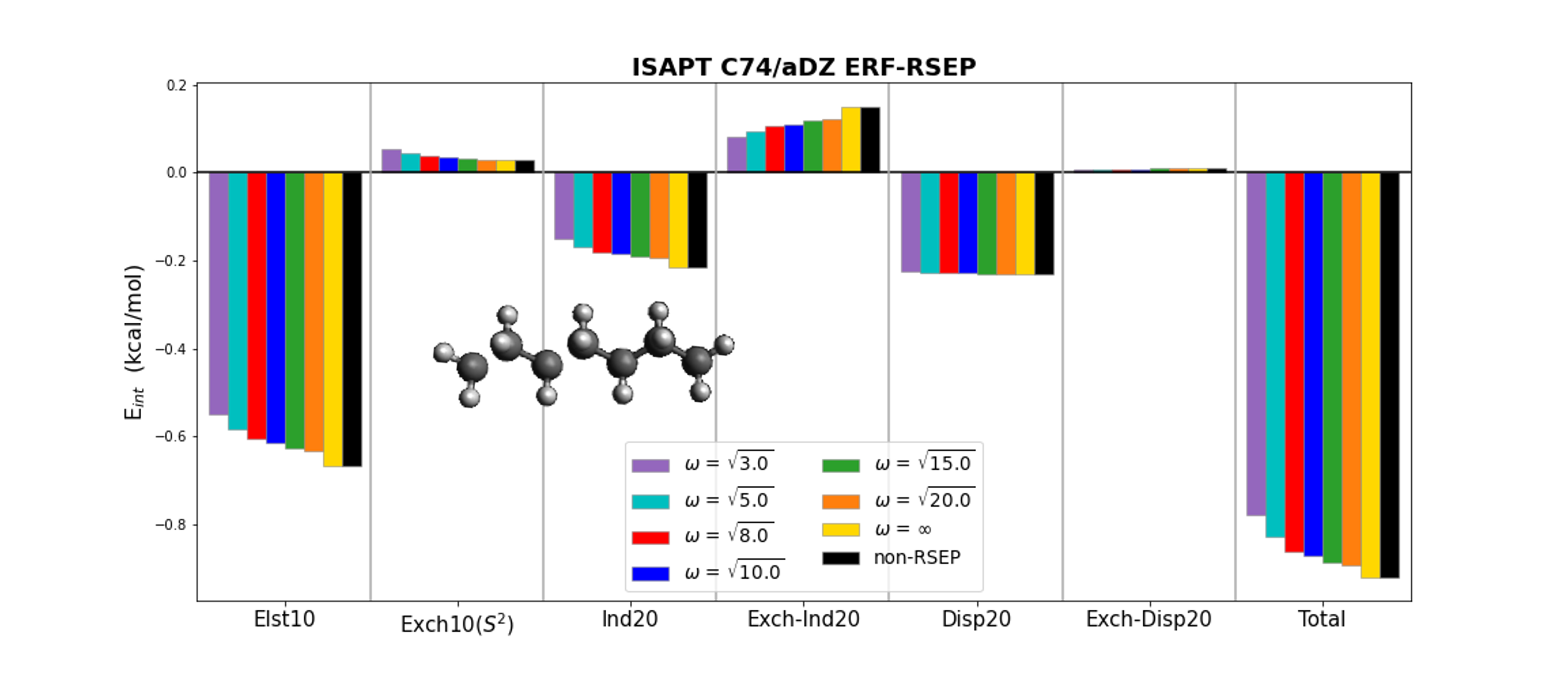}
\end{minipage}
\vspace{-.01in}
\begin{minipage}{.48\textwidth}
\centering
\includegraphics[width=1\textwidth]{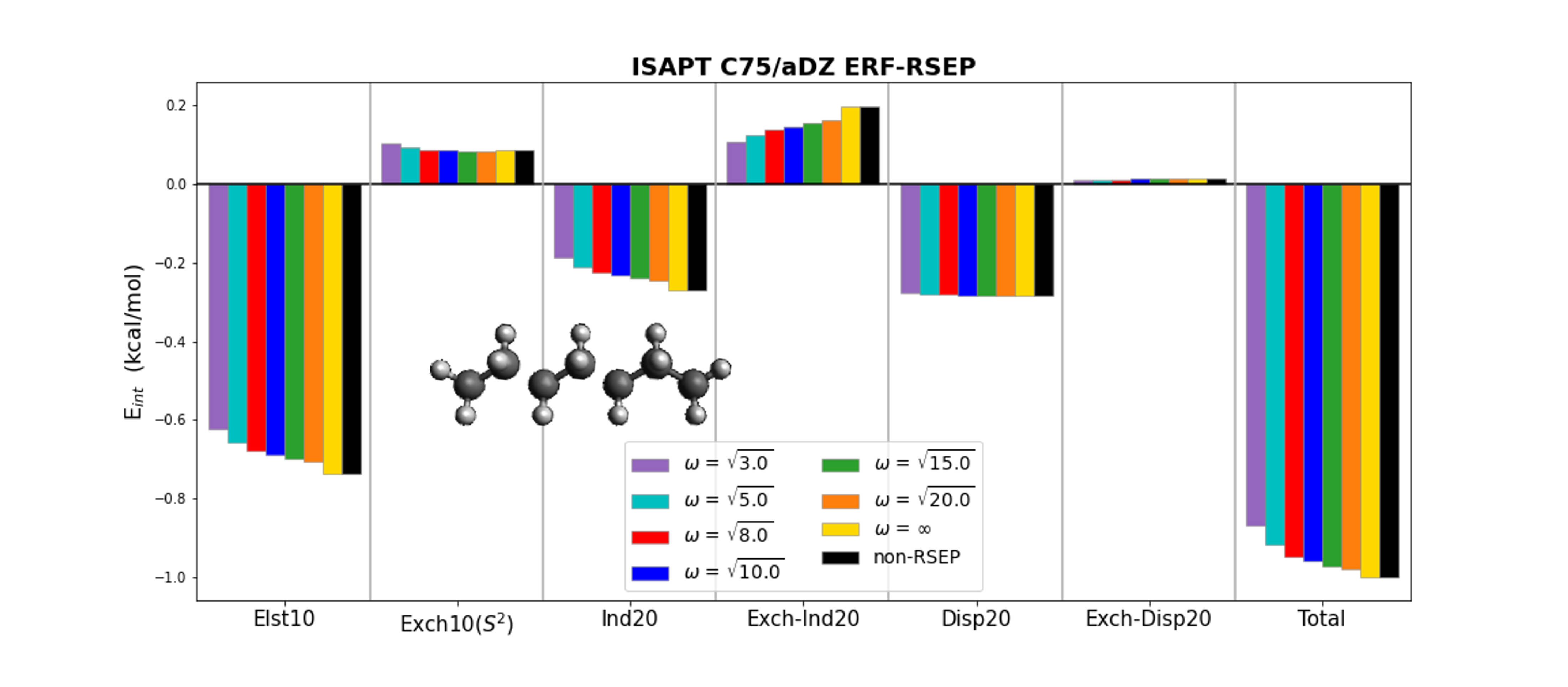}
\end{minipage}
\vspace{-.21in}
\begin{minipage}{.48\textwidth}
\centering
\includegraphics[width=1\textwidth]{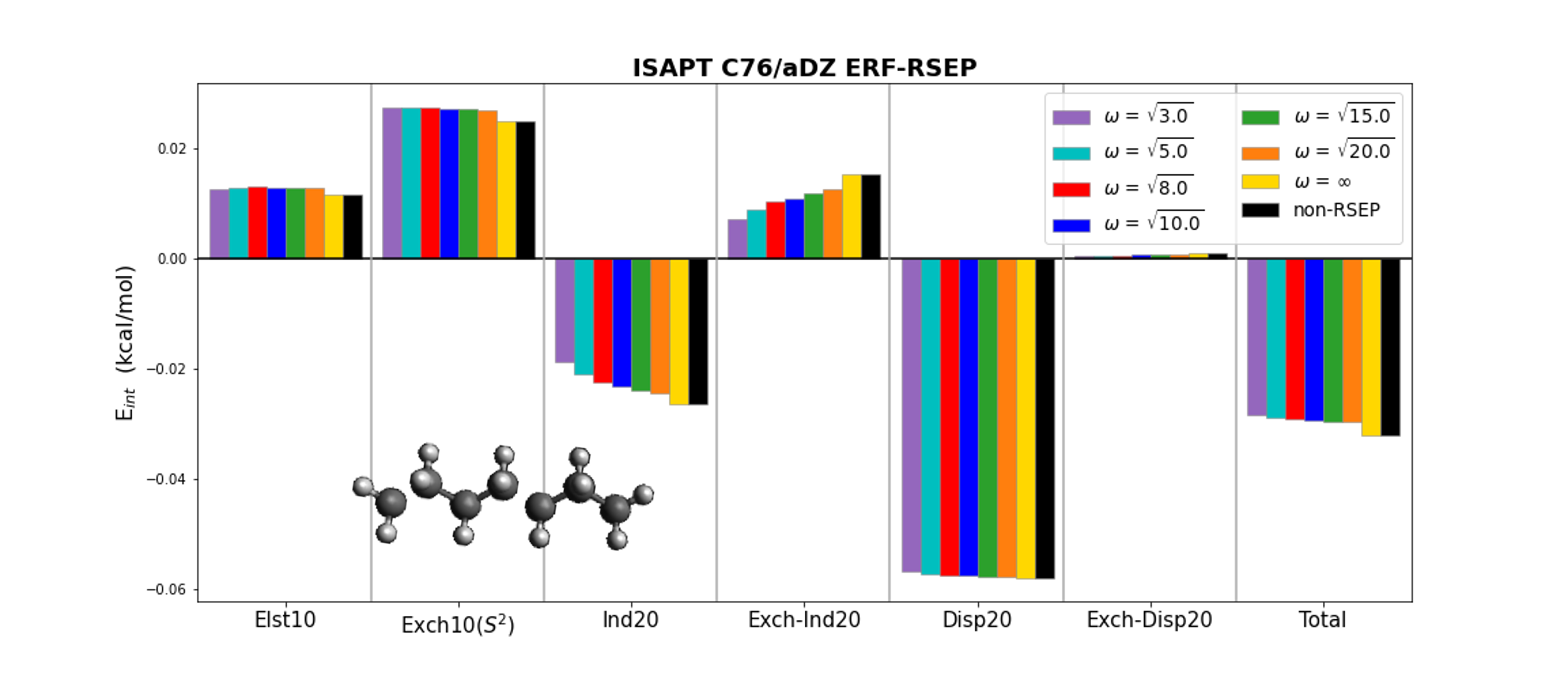}
\end{minipage}
\vspace{-.21in}
\begin{minipage}{.48\textwidth}
\centering
\includegraphics[width=1\textwidth]{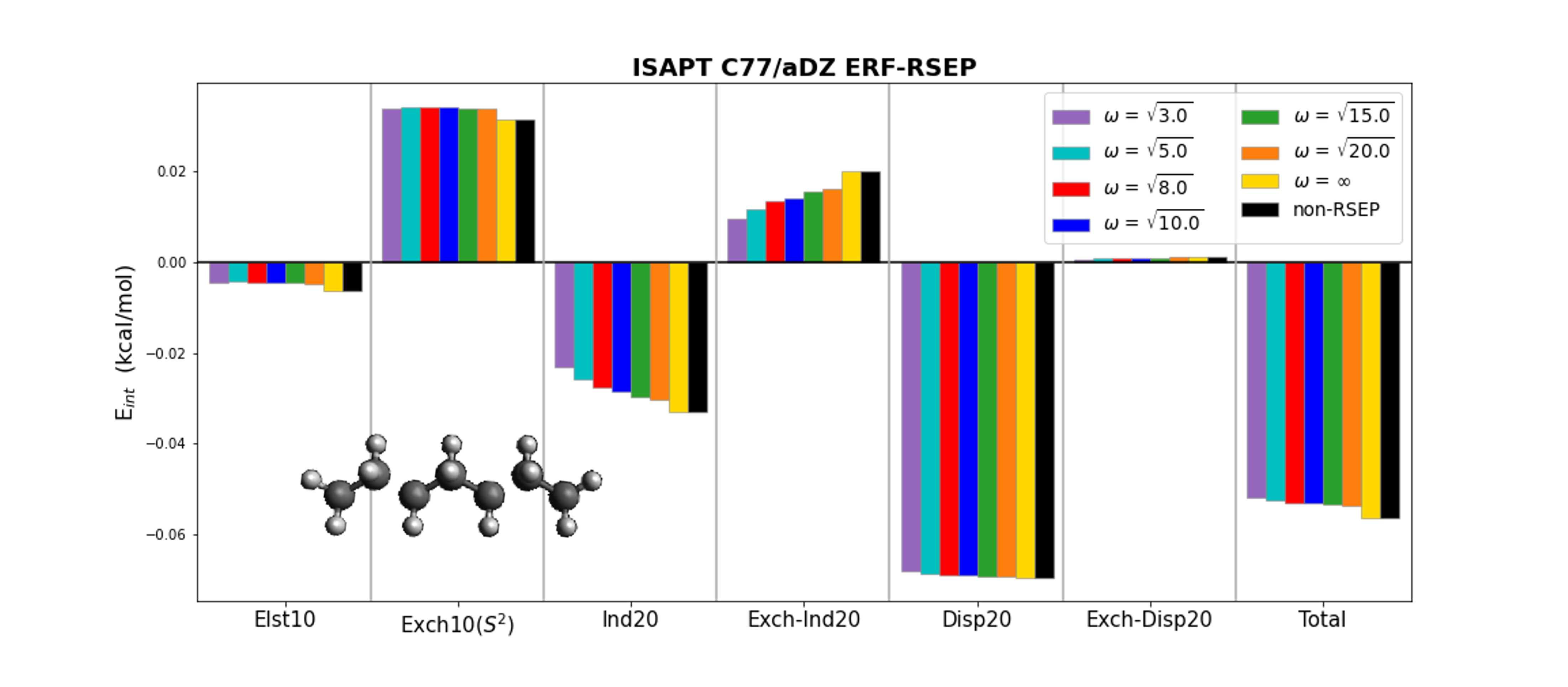}
\end{minipage}
\vspace{-.02in}
\begin{minipage}{.48\textwidth}
\centering
\includegraphics[width=1\textwidth]{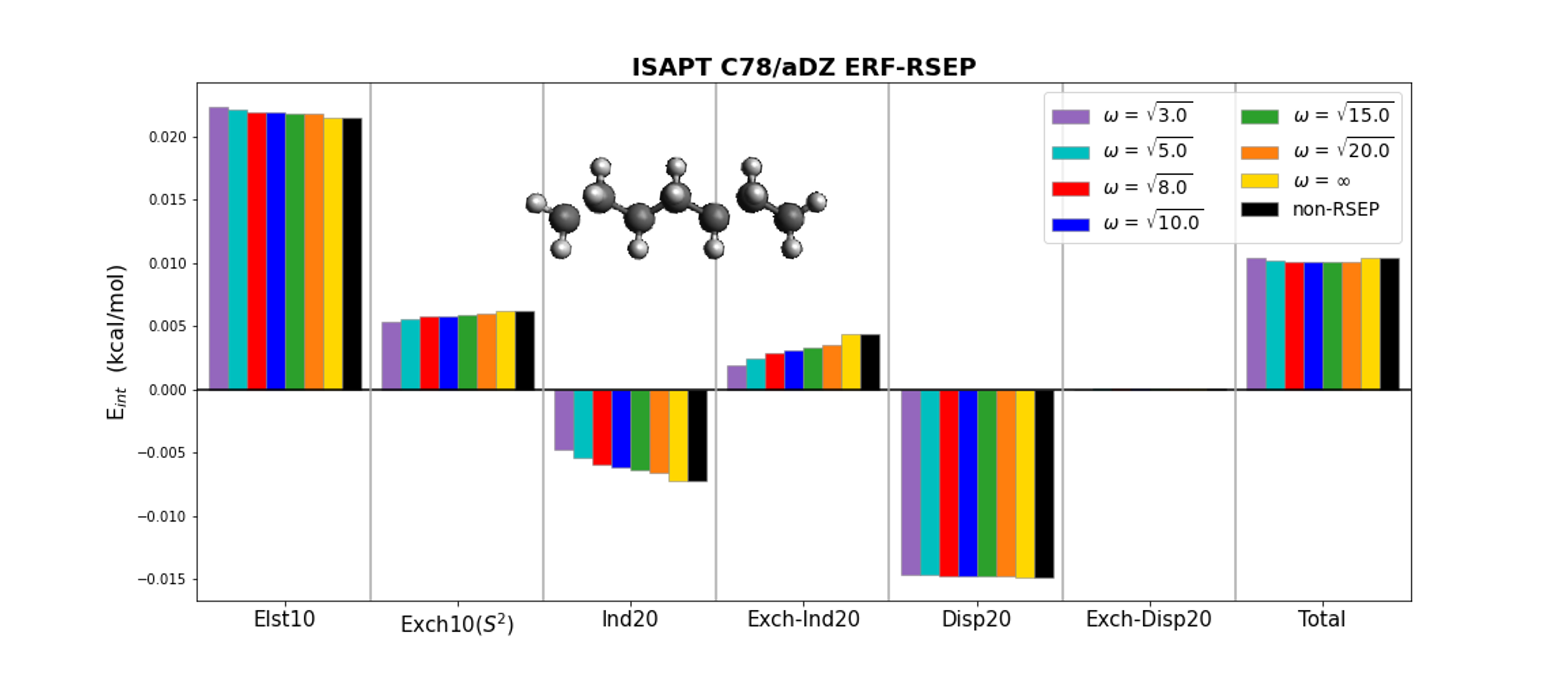}
\end{minipage}
\vspace{-.02in}
\begin{minipage}{.48\textwidth}
\centering
\includegraphics[width=1\textwidth]{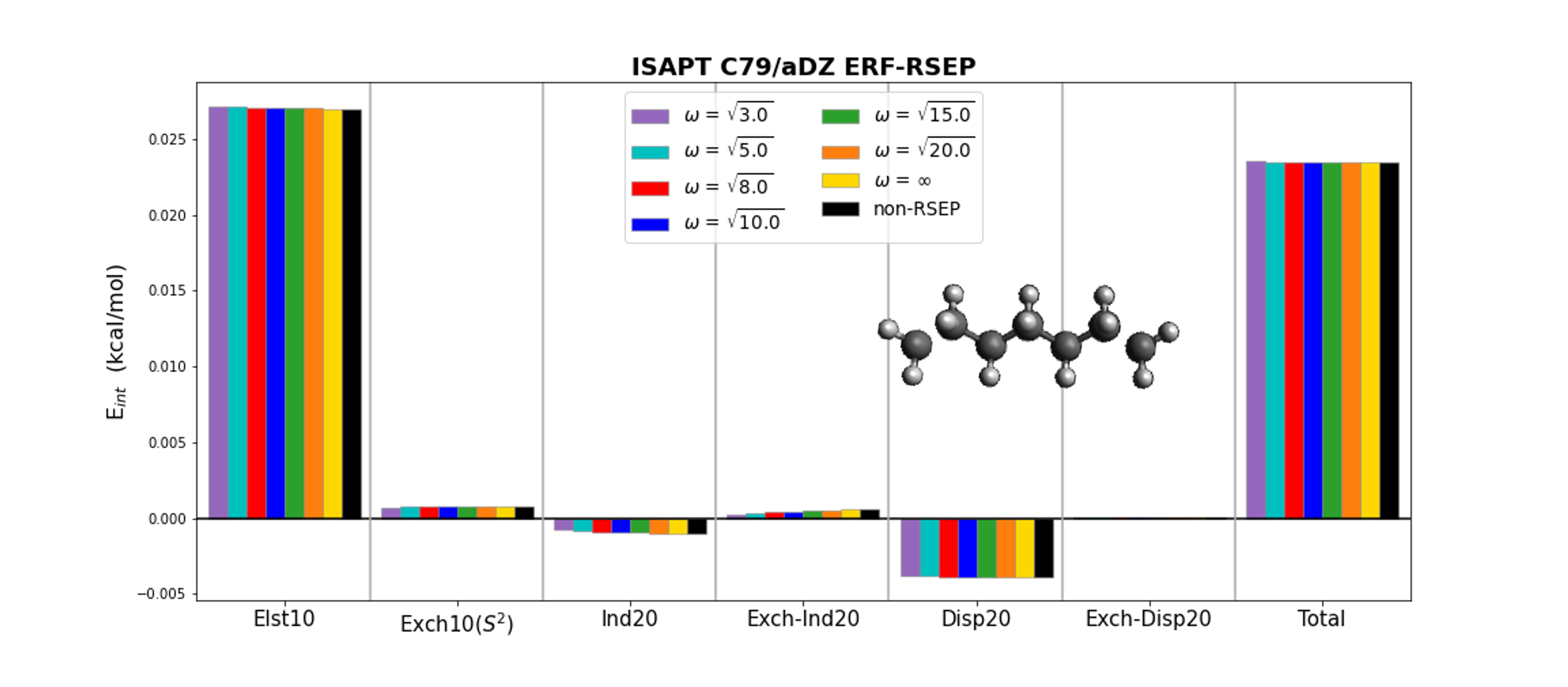}
\end{minipage}
\caption{The components of the ISAPT/SIAO1 (RSEP/erf) interaction energy for 9 fragmentation models (pictured) of the n-heptane molecule computed with different values of $\omega$ (bohr$^{-1}$) in the aDZ basis set. The energy results are compared to those of the traditional ISAPT/SIAO1 method.}
\label{fig:c7h16-erf-memdf-plots}
\end{figure}

The consistency of the range-separated SAPT0 interaction energy components with varied $\eta/\omega$ is quite promising. 
As the next step and an even more stringent test, we decided to examine the influence of the purely long-ranged interaction potential $V_{\rm 1e2e}$, in the RSEP/gau and RSEP/erf versions, on the accuracy of the ISAPT/SIAO1 intramolecular scheme. 
The long-ranged one-electron and two-electron potential integrals are applied throughout the calculation of the energy corrections. 
We investigate a large subset of intramolecular interactions studied in Ref.~\citenum{Luu:23}, including multiple fragmentations of the family of pentanediol (1,4-; 1,5-; 2,4-) molecules as well as $n$-heptane.
The resulting convergence of ISAPT/SIAO1 energy contributions as functions of $\eta$ ($\omega$) is presented in Figs. \ref{fig:P141-adz-gau-erf-memdf-plot}--\ref{fig:c7h16-erf-memdf-plots}. 
Note that the interaction energies for different fragmentation patterns are very different from each other: this is expected as some fragments are closer and others are relatively far away, separated by a large linker.

To perform range-separated intramolecular calculations, we implemented the RSEP/gau and RSEP/erf algorithms into the ISAPT/SIAO1 method in the {\sc psi4} code \cite{Smith:20} with density fitting.
The most complex part of the implementation was the calculation of long-range generalized Coulomb and exchange (JK) matrices which are pervasive in atomic orbital-basis SAPT expressions.
To verify the correctness of our code, we implemented the RSEP algorithm into two density-fitted JK algorithms present in {\sc psi4}, the memory-based one (memDF) and the disk-based one (diskDF).
The data in Figs. \ref{fig:P141-adz-gau-erf-memdf-plot}--\ref{fig:c7h16-erf-memdf-plots}, and for other fragmentation models presented in the Supporting Information, were obtained with the memDF algorithm. 
We repeated all of the same calculations using an alternative diskDF algorithm and confirmed that all ISAPT energies are the same (the absolute difference in total ISAPT values does not exceed $2.09\times 10^{-6}$ kcal/mol).
It should be stressed that, while we used the aDZ orbital basis in all ISAPT calculations, the default
aDZ-RI choice for the DF basis used to expand dispersion amplitudes was not always adequate and sometimes led to erratic dispersion energies as a function of $\eta$. Therefore, all range-separated ISAPT/SIAO1 results presented here have been obtained with a larger aTZ-RI dispersion auxiliary basis. 
The fragmentation pattern names for all systems, given in the header of each figure, follow the definitions of Ref.~\citenum{Luu:23}; in addition, we display the fragmentation in each figure as a ball-and-stick model with the two interfragment bonds removed.

%\clearpage

%\clearpage

First, we verified that the ISAPT/SIAO1 (RSEP/gau) and ISAPT/SIAO1 (RSEP/erf)  interaction energies exhibit the correct limiting behavior for $\eta/\omega\to 0$ and $\eta/\omega\to\infty$. 
The energy corrections collapse to zero with zeroed long-ranged interaction potentials when $\eta/\omega$ = 0.0, and accurately recover the non-range separated ISAPT/SIAO1 data when $\eta$ $(\omega)$ is on the order of 10000000 bohr$^{-2}$ (10000000 bohr$^{-1}$) (shown as $\eta/\omega$ = $\infty$ on energy graphs, Figs. \ref{fig:P141-adz-gau-erf-memdf-plot}-\ref{fig:c7h16-erf-memdf-plots}). 
The interaction energy components exhibit reasonable convergence towards the conventional ISAPT/SIAO1 results, with a somewhat smoother and quicker convergence  observed in the RSEP/erf variant.
In the ISAPT/SIAO1 (RSEP/gau) method, the calculations using $\eta\leq 5$ bohr$^{-2}$ may overestimate or flip the sign of the electrostatic energy due to the change of the Coulomb potential being too drastic. 
The total interaction energy is affected similarly, and its changes with $\eta$ typically follow the changes in the electrostatic energy. 
On the other hand, the ISAPT/SIAO1 (RSEP/erf)  electrostatic energy exhibits less variance in  $\omega$ for all tested models, but still exhibits some small discrepancies when $\omega <$ 5 bohr$^{-1}$. 
Notably, the variance in the electrostatic energy resulting from the small values of $\eta/\omega$ is unrelated to the size of the linker fragment in all tested models. 
For example, the linker fragment of the P244 model is -CH$_2$-; consequently, the occupied orbitals of two fragments C$_2$H$_4$OH-, which are embedded in the Coulomb and exchange operators of the fragment linker, participate in the calculation of all energy corrections at the SAPT0 level. 
In contrast, the original ISAPT/C linker partitioning algorithm \cite{Parrish:15} works mostly well for large linker fragments but fails for small linkers \cite{Luu:23}.

For the Gaussian splitting with $\eta\geq 8$ bohr$^{-2}$ and the error-function splitting with $\omega\geq\sqrt{5}$ bohr$^{-1}$, a convergent pattern emerges and the electrostatic energies are in qualitative agreement with the conventional values, even though the convergence of the small residual short-range effects can be quite slow. 
At large $\eta$, the difference between $1/r$ and $v_{\rm l}(r)$ is nonnegligible only at very small $r$, so the residual effects missing in the long-range electrostatic energy are the consequence of {\bf (i)} nonzero electron density of one fragment at the location of the other fragment's nuclei, and {\bf (ii)} nonzero probability of finding a pair of electrons from different fragments at the same location in space.
We observed that the effects {\bf (i)} and {\bf (ii)} are of the same order of magnitude. 
The large-$\eta$ and large-$\omega$ convergence in the 1,4-pentanediol case and the P141 fragmentation is illustrated in Fig.~\ref{fig:P141-adz-gau-erf-memdf-plot}  for the aDZ basis, and in the Supporting Information for the aTZ and aQZ ones.
In this case, the effect {\bf (i)} dominates, resulting in a missing attractive nucleus-electron contribution at finite $\eta$ and the electrostatic energy converging to its $\eta\to\infty$ limit from above.  
For this system, we further broke down the electrostatic interaction energy into its nuclear-electron and electron-electron contributions, illustrating the $\eta/\omega$ dependence of each term -- see the Supporting Information for the relevant figures (the long-range nuclear repulsion term is already converged to the non-range separated value by $\eta=10$ bohr$^{-2}$). 
Thus, the slow convergence of the small residual terms in the electrostatic energy is a consequence of the incomplete cancellation between the nuclear-electron and electron-electron residual electrostatic energies, where either effect may prevail. 
In the comparison to the Gaussian splitting (colored bars in Fig.~\ref{fig:P141-adz-gau-erf-memdf-plot}), the error-function one (empty bars) produces faster convergence of the electrostatic energy at large $\omega$, with some cancellation between the residual one- and two-electron effects taking place as well (see the Supporting Information for the relevant figures).

Going beyond the electrostatic energy, the convergence of other ISAPT corrections to their non-range separated values is mostly quick, monotonic, and systematic. 
The induction and exchange-induction terms may slightly underestimate their conventional values even at $\eta=20$ bohr$^{-2}$, although the missing residual effects cancel to a large extent between $E^{(20)}_{\rm ind,resp}$ and $E^{(20)}_{\rm exch-ind,resp}$.
The convergence of dispersion energy with $\eta/\omega$ is notably quick and systematic. 
The first-order exchange term also converges monotonously to its limit, however, a small residual contribution is occasionally still missing at $\eta=20$ bohr$^{-2}$. 
However, all non-electrostatic terms vary with the range separation parameter less than the electrostatic energy, so the convergence of the total ISAPT interaction energy to its non-range separated value closely follows the convergence pattern of $E^{(10)}_{\rm elst}$. 
In all tested systems, when $\omega$ is equivalent to the $\eta$ value, e.g. $\eta$ = 20.0 bohr$^{-2}$ versus $\omega$ = $\sqrt{20.0}$ bohr$^{-1}$, the induction and exchange induction energies have about 4.6\% and 8.9\% smaller average errors, respectively, when using the error function.
%{\bf (KP: This last sentence is very interesting but needs to be made clearer and more precise. Could we use a pair where $\eta=\omega^2$ exactly? Also, it's not clear what a ``20\%\ quicker convergence'' means, it's better to say something like ``20\%\ smaller errors''.)}
Overall, we observe that the neglect of the short-range Coulomb potential still leads to reasonable intramolecular SAPT energy contributions, especially within the error-function range separation. 
On the other hand, the Gaussian range separation used in the ISAPT/SIAO1 method may prefer larger values of the parameter, $\eta\geq  8$ bohr$^{-2}$, than for the analogous intermolecular range-separated SAPT approach.

\section{Conclusions}

Symmetry-adapted perturbation theory is commonly used to compute and decompose noncovalent interaction energies between two molecules or molecular fragments. 
SAPT does provide both a long-range multipole expansion of interaction energy and short-range, overlap-dependent effects.
An alternative separation between long-range and short-range interaction, common in the DFT context, is via an explicit splitting of the Coulomb operator $1/r$ into its long-range ($v_{\rm l}(r)$) and short-range ($v_{\rm s}(r)$) parts.
Such a separation has been used in SAPT as well, but only in the specific contexts of preventing divergence of the perturbation series for a model system \cite{Patkowski:04} and separating induction effects into polarization and charge transfer \cite{Misquitta:13}.
In this work, we follow a different direction and investigate the behavior of intermolecular (SAPT0) and intramolecular (ISAPT) energy terms when the full Coulomb interaction operator is approximated by its long-range part.
In a sense, we build on the study of Chen et al. \cite{Chen:16} who performed a similar approximation in various supermolecular interaction energies for rare gas dimers, but we are able to examine each SAPT and ISAPT contribution separately.
Two previously proposed forms of splitting the Coulomb operator are employed, a Gaussian range separation and an error-function one.

Our main observation is that the long-range part of the interaction potential $v_l(r)$ is sufficient for a faithful description of noncovalent interactions.
This result could not have been taken for granted and should be contrasted with the insufficiency of the long-range multipole expansion, which completely misses both charge penetration contributions and all exchange terms.
In our case, the first-order range-separated electrostatic and exchange energies for model intermolecular complexes exhibit smooth convergence to the full non-range separated energies when the separation parameter $\eta/\omega$ goes to infinity.
Moreover, the first-order terms with range separation are close to the nonapproximated values at near-minimum and larger intermolecular separations as long as  $\eta \geq 3.0$ bohr$^{-2}$ or $\omega \geq \sqrt{3.0}$ bohr$^{-1}$.
It is also notable that the full range separation RSEP overperforms its one-electron-only variant RSEP-eN for most systems, indicating that the approximation can be made more severe (neglecting more terms) as long as the attractive and repulsive contributions are treated in a balanced way.
Similarly, the RSEP second-order induction and exchange-induction terms are highly consistent between different values of $\eta/\omega$, much more so than the RSEP-eN ones. This often, but not always, translates to a more accurate description of the sum $E^{(20)}_{\rm ind,resp}+E^{(20)}_{\rm exch-ind,resp}$ as well.
The dispersion and exchange dispersion energies exhibit virtually no dependence on $\eta/\omega$ in the RSEP-eN variant (as the two-electron integrals are unchanged) and moderate $\eta/\omega$ dependence for RSEP.
As far as total SAPT0(RSEP) energies are concerned, the differences in first-order electrostatic and exchange terms, while partially canceling each other, dominate the residual error of the range-separated approximation; this error is small at the van der Waals minimum and larger separations.
Overall, the full RSEP scheme using the error function splitting provides the best agreement with conventional SAPT0.

The application of range separation to intramolecular SAPT (in the SIAO1 linker assignment scheme of Ref.~\citenum{Luu:23}) shows similar smooth and mostly monotonic convergence to the non-range separated values, but due to a stronger interaction, the requirements for the range separation parameter are more strict.
The $\eta\leq 5$ bohr$^{-2}$ choice, representing a very crude approximation in the ISAPT case, often overestimates or flips the sign of the electrostatic energy. 
The total interaction energy is affected similarly, as electrostatic energy is typically the leading contributor to the ISAPT energy changes with $\eta$; all other first- and second-order terms converge faster.
Overall, the $\eta\geq 8$ bohr$^{-2}$ values afford reasonably accurate range-separated ISAPT contributions, although small residual errors remain even for $\eta=20$ bohr$^{-2}$. 
The error-function range separation provides superior consistency and fairly rapid convergence in all interaction energy components. The relatively small energy variance between $\omega$ values illustrates that the long-ranged error-function potential is the optimal selection for the range separation-based ISAPT method.  

The results of this work demonstrate that all noncovalent interaction energy contributions can be well described by only the long-range part of the Coulomb potential. 
When the interaction gets particularly strong, such as for intermolecular complexes at repulsive short-range separations or for some intramolecular complexes, the requirements on the range separation parameter become stricter, but the purely long-range treatment remains accurate at least qualitatively, and often quantitatively.
This suggests that one may construct an alternative ISAPT algorithm that replaces molecular fragmentation by range separation of the Coulomb potential, yielding the entire nonbonded energy within a molecule rather than an interaction energy between two specific fragments. Work in this direction is in progress in our group.

\section*{Associated Content}

\noindent {\bf Supporting Information:} The Supporting Information is available free of charge at xxx/xxx/xxx.

Figures displaying range-separated SAPT0 and ISAPT data for additional systems and basis sets.

\section*{Acknowledgments}

D.L. and K.P. are supported by the U.S. National Science Foundation award CHE-1955328.
C.C. and K.P. thank the Swiss National Science Foundation (Grant IZSEZ0\_180167) for support.

\section*{Appendix: Range-Separated One-Electron Integrals}

%For now, Du's note was copied and pasted here without changes.

The standard one-electron auxiliary nuclear attraction integral:

\begin{equation}
    \langle \mathbf{0}_A|\frac{1}{r}| \mathbf{0}_B \rangle ^{(m)} = \frac{2}{\sqrt{\pi}}\zeta ^ {(1/2)} (\mathbf{0}_A||\mathbf{0}_B)\mathbf{F}_m(T)
\end{equation}
where $(\mathbf{0}_{A}||\mathbf{0}_{B})$ is the overlap integral of two s-type Gaussian centered at $\mathbf{A}$ and $\mathbf{B}$, $\zeta = \zeta_{a} + \zeta_{b}$ is the Gaussian exponent, and $\mathbf{F}_{m}(T)$ is the Boys function to evaluate Coulomb integrals.

The Boys function of order m is defined by:

\begin{equation}
    \mathbf{F}_{m}(T) = \int_{0}^{1} \text{exp}(-Tt^2)t^{2m}dt
\end{equation}

and its derivatives:
\begin{equation}
    \frac{\text{d}\mathbf{F}_{m}(T)}{\text{d}T} = -\mathbf{F}_{m+1}(T)
\end{equation}

The auxiliary integral using the Boys function $\mathbf{F}_{m}$ in a recursive scheme:

\begin{align}
\mathbf{R}^m_{tuv} (p, \mathbf{R}_{PC}) = (\frac{\partial}{\partial P_x})^t (\frac{\partial}{\partial P_y})^u (\frac{\partial}{\partial P_z})^v \mathbf{R}^m_{000}(p, \mathbf{R}_{PC}) \\
\mathbf{R}^m_{000}(p, \mathbf{R}_{PC}) = (-2p)^m \mathbf{F}_m (pR_{PC}^2) \\
R_{PC}^2 = (P_x - C_x)^2 + (P_y - C_y)^2 + (P_z - C_z)^2
\end{align}

where $t, u, v$ are the angular momenta of the two-center overlap Gaussian at point $\mathbf{P}$ = $\frac{\zeta_a \mathbf{A} + \zeta_b \mathbf{B}}{\zeta_a + \zeta_b}$, with the Gaussian exponent $p = p_a + p_b$, for a nucleus located at $\mathbf{C}$.

\subsection*{The Gaussian basis function}

The splitting of the Coulomb potential into long-range $v_l(s)$ and short-range $v_s(s)$ using the Gaussian function can be expressed as:
\begin{align}
    \frac{1}{r} = v_l(s) + v_s(s) \\ 
    v_l(s) = \frac{1-e^{-\eta r^2}}{r} \\  
    v_s(s) = \frac{e^{-\eta r^2}}{r}
\end{align}

For $m = 0$, the s-type nuclear attraction integral using the $v_s(r)$ potential at the 0\textit{th} order of the Boys function is:

\begin{align}
\mathbf{R}^0_{000} (\frac{p^2}{p + \eta}, \mathbf{R}_{PC}) = \langle \mathbf{0}_A|\frac{1}{r}| \mathbf{0}_B \rangle ^{(0)}_{\eta} = \frac{p}{p + \eta} \text{exp}(\frac{-p\eta}{p + \eta} R_{PC}^2) \mathbf{F}_0 (\frac{p^2}{p + \eta} R_{PC}^2) 
\end{align}

Incrementing over $t$ :

\begin{align}
\mathbf{R}^0_{100} (\frac{p^2}{p + \eta}, \mathbf{R}_{PC}) = & \frac{\partial}{\partial P_x} \mathbf{R}^0_{000} (\frac{p^2}{p + \eta}, \mathbf{R}_{PC}) \nonumber \\ &
= \frac{p}{p + \eta} \mathbf{F}_0 (\frac{p^2}{p + \eta} R_{PC}^2) \text{exp}(\frac{-p\eta}{p + \eta} \mathbf{R}_{PC}^2) (\frac{-p\eta}{p + \eta})(2P_x - 2C_x) \nonumber \\ &
+ \frac{p}{p + \eta} \text{exp}(\frac{-p\eta}{p + \eta} \mathbf{R}_{PC}^2) (-\mathbf{F}_1 (\frac{p^2}{p + \eta} R_{PC}^2))\frac{p^2}{p + \eta}(2P_x - 2C_x) \nonumber \\ &
= (\frac{-2p\eta}{p + \eta})\mathbf{R}^0_{000}(\frac{p^2}{p + \eta}, \mathbf{R}_{PC})\mathbf{X}_{PC} + \mathbf{R}^1_{000}(\frac{p^2}{p + \eta}, \mathbf{R}_{PC})\mathbf{X}_{PC} \nonumber \\ &
= \mathbf{X}_{PC}[(\frac{-2p\eta}{p + \eta})\mathbf{R}^0_{000}(\frac{p^2}{p + \eta}, \mathbf{R}_{PC}) + \mathbf{R}^1_{000}(\frac{p^2}{p + \eta}, \mathbf{R}_{PC}) ]
\end{align}

\begin{align}
\mathbf{R}^0_{t+1,0,0} (\frac{p^2}{p + \eta}, \mathbf{R}_{PC}) = & (\frac{\partial}{\partial P_x})^t \frac{\partial}{\partial P_x}\mathbf{R}^0_{000} (\frac{p^2}{p + \eta}, \mathbf{R}_{PC}) = (\frac{\partial}{\partial P_x})^t\mathbf{R}^0_{100} (\frac{p^2}{p + \eta}, \mathbf{R}_{PC}) \nonumber \\ &
= (\frac{\partial}{\partial P_x})^t [\mathbf{X}_{PC}(\frac{-2p\eta}{p + \eta}\mathbf{R}^0_{000} + \mathbf{R}^1_{000} )] \nonumber \\ &
= t(\frac{\partial}{\partial P_x})^{t-1}(\frac{-2p\eta}{p + \eta}\mathbf{R}^0_{000} + \mathbf{R}^1_{000}) + \mathbf{X}_{PC}(\frac{\partial}{\partial P_x})^t(\frac{-2p\eta}{p + \eta}\mathbf{R}^0_{000} + \mathbf{R}^1_{000}) \nonumber \\ &
= t(\frac{-2p\eta}{p + \eta})\mathbf{R}^0_{t-1,0,0} + t\mathbf{R}^1_{t-1,0,0} + \mathbf{X}_{PC}(\frac{-2p\eta}{p + \eta})\mathbf{R}^0_{t,0,0} + \mathbf{X}_{PC}\mathbf{R}^1_{t,0,0}
\end{align}

Applying the angular recursion relation over $t, u, v$:

\begin{align}
\mathbf{R}^n_{t+1,u,v} (\frac{p^2}{p + \eta}, \mathbf{R}_{PC}) = t(\frac{-2p\eta}{p + \eta})\mathbf{R}^n_{t-1,u,v} + t\mathbf{R}^{n+1}_{t-1,u,v} + \mathbf{X}_{PC}(\frac{-2p\eta}{p + \eta})\mathbf{R}^n_{tuv} + \mathbf{X}_{PC}\mathbf{R}^{n+1}_{tuv} \\ 
\mathbf{R}^n_{t,u+1,v} (\frac{p^2}{p + \eta}, \mathbf{R}_{PC}) = u(\frac{-2p\eta}{p + \eta})\mathbf{R}^n_{t,u-1,v} + u\mathbf{R}^{n+1}_{t,u-1,v} + \mathbf{Y}_{PC}(\frac{-2p\eta}{p + \eta})\mathbf{R}^n_{tuv} + \mathbf{Y}_{PC}\mathbf{R}^{n+1}_{tuv} \\
\mathbf{R}^n_{t,u,v+1} (\frac{p^2}{p + \eta}, \mathbf{R}_{PC}) = v(\frac{-2p\eta}{p + \eta})\mathbf{R}^n_{t,u,v-1} + v\mathbf{R}^{n+1}_{t,u,v-1} + \mathbf{Z}_{PC}(\frac{-2p\eta}{p + \eta})\mathbf{R}^n_{tuv} + \mathbf{Z}_{PC}\mathbf{R}^{n+1}_{tuv} 
\end{align}

The auxiliary nuclear attraction integral in n\textit{th} order of the Boys function:

\begin{align}
\mathbf{R}^n_{000} (\frac{p^2}{p + \eta}, \mathbf{R}_{PC}) = \frac{p}{p + \eta} \text{exp}(\frac{-p\eta}{p + \eta} R_{PC}^2) (\frac{-2p^2}{p + \eta})^n \mathbf{F}_n (\frac{p^2}{p + \eta} R_{PC}^2) 
\end{align}

\subsection*{The error function}

The splitting of the Coulomb potential into long-range $v_l(s)$ and short-range $v_s(s)$ using the standard error function can be expressed as:
\begin{align}
    \frac{1}{r} = v_l(r) + v_s(r) \\ 
    v_l(r) = \frac{\text{erf}(\omega r)}{r} \\  
    v_s(r) = \frac{1-\text{erf}(\omega r)}{r}
\end{align}

For $m = 0$, the s-type nuclear attraction integral using the $v_l(r)$ potential at the 0\textit{th} order of the Boys function is:

\begin{align}
\mathbf{R}^0_{000} (\frac{p\omega^2}{p + \omega^2}, \mathbf{R}_{PC}) = \langle \mathbf{0}_A|\frac{1}{r}| \mathbf{0}_B \rangle ^{(0)}_{\omega} = (\frac{\omega^2}{p + \omega^2})^{1/2}  \mathbf{F}_0 (\frac{p\omega^2}{p + \omega^2}R_{PC}^2) 
\end{align}

Incrementing over $t$ :

\begin{align}
\mathbf{R}^0_{100} (\frac{p\omega^2}{p + \omega^2}, \mathbf{R}_{PC}) = & \frac{\partial}{\partial P_x} \mathbf{R}^0_{000} (\frac{p\omega^2}{p + \omega^2}, \mathbf{R}_{PC}) \nonumber \\ &
= (\frac{\omega^2}{p + \omega^2})^{1/2} (-\mathbf{F}_1 (\frac{p\omega^2}{p + \omega^2}R_{PC}^2))\frac{p\omega^2}{p + \omega^2}(2P_x - 2C_x) \nonumber \\ &
= \mathbf{X}_{PC}\mathbf{R}^1_{000}(\frac{p\omega^2}{p + \omega^2}, \mathbf{R}_{PC})
\end{align}

\begin{align}
\mathbf{R}^0_{t+1,0,0} (\frac{p\omega^2}{p + \omega^2}, \mathbf{R}_{PC}) = & (\frac{\partial}{\partial P_x})^t \frac{\partial}{\partial P_x}\mathbf{R}^0_{000} (\frac{p\omega^2}{p + \omega^2}, \mathbf{R}_{PC}) = (\frac{\partial}{\partial P_x})^t\mathbf{R}^0_{100} (\frac{p\omega^2}{p + \omega^2}, \mathbf{R}_{PC}) \nonumber \\ &
= (\frac{\partial}{\partial P_x})^t \mathbf{X}_{PC}\mathbf{R}^1_{000}  \nonumber \\ &
= t(\frac{\partial}{\partial P_x})^{t-1}\mathbf{R}^1_{000} + \mathbf{X}_{PC}(\frac{\partial}{\partial P_x})^t \mathbf{R}^1_{000} \nonumber  \nonumber \\ &
= t\mathbf{R}^1_{t-1,0,0} + \mathbf{X}_{PC}\mathbf{R}^1_{t,0,0}
\end{align}

Applying the angular recursion relation over $t, u, v$:

\begin{align}
\mathbf{R}^n_{t+1,u,v} (\frac{p\omega^2}{p + \omega^2}, \mathbf{R}_{PC}) = t\mathbf{R}^{n+1}_{t-1,u,v} + \mathbf{X}_{PC}\mathbf{R}^{n+1}_{tuv} \\ 
\mathbf{R}^n_{t,u+1,v} (\frac{p\omega^2}{p + \omega^2}, \mathbf{R}_{PC}) = u\mathbf{R}^{n+1}_{t,u-1,v} + \mathbf{Y}_{PC}\mathbf{R}^{n+1}_{tuv} \\
\mathbf{R}^n_{t,u,v+1} (\frac{p\omega^2}{p + \omega^2}, \mathbf{R}_{PC}) = v\mathbf{R}^{n+1}_{t,u,v-1} + \mathbf{Z}_{PC}\mathbf{R}^{n+1}_{tuv} 
\end{align}

The auxiliary nuclear attraction integral in n\textit{th} order of the Boys function:

\begin{align}
\mathbf{R}^n_{000} (\frac{p\omega^2}{p + \omega^2}, \mathbf{R}_{PC}) = (\frac{\omega^2}{p + \omega^2})^{1/2} (\frac{-2p\omega^2}{p + \omega^2})^n \mathbf{F}_n (\frac{p\omega^2}{p + \omega^2}R_{PC}^2) 
\end{align}

\clearpage

\bibliography{master}

\clearpage

\begin{figure}
\caption{Graphical Table of Contents Image}
\begin{center}
\includegraphics{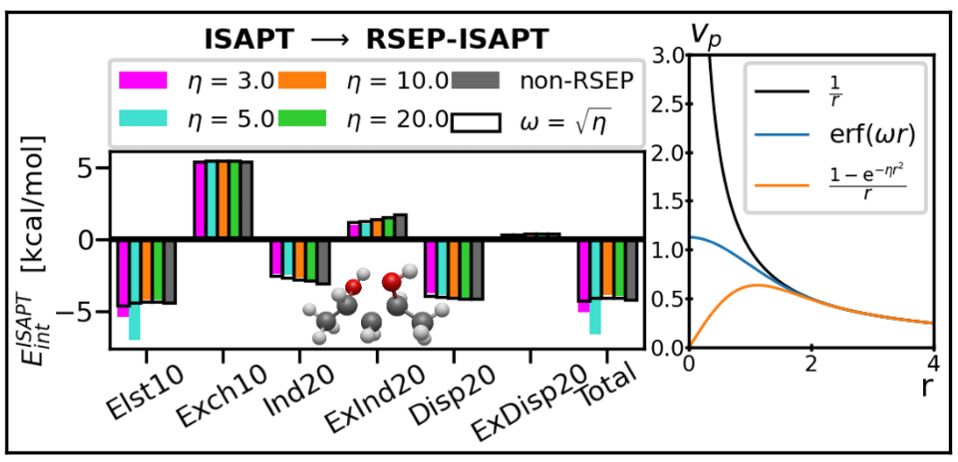}
\end{center}
\end{figure}

\end{document}